\pdfoutput=1

\documentclass[longauth]{config/aa}  

\usepackage{natbib}
\usepackage{comment}
\usepackage{graphicx}
\usepackage{multirow}
\usepackage{booktabs}
\usepackage{textgreek}
\usepackage{xargs}
\usepackage[dvipsnames]{xcolor}
\usepackage{xspace}
\usepackage{caption}
\usepackage{float}
\usepackage{enumitem}
\usepackage{siunitx}

\usepackage{txfonts}

\usepackage{subcaption}
\usepackage{adjustbox}

\newcommand{\logU}{\ensuremath{\log U}}

\graphicspath{{./}{figs/}}
\usepackage{etoolbox}
\makeatletter
\newcommand\sendemail[3]{
\edef\@tempa{mailto:#1?subject=#2 }%
\edef\@tempb{\expandafter\html@spaces\@tempa\@empty}%
\href{\@tempb}{#3}}

\catcode\%=11
\def\html@spaces#1 #2{#1
\catcode\%=14
\makeatother

\newcommand{\citationneeded}{\textcolor{ForestGreen}{$^{\rm citation\;needed}$}}
\let\oldtextsigma\textsigma
\renewcommand{\textsigma}{\oldtextsigma\xspace}
\let\oldtextalpha\textalpha
\renewcommand{\textalpha}{\oldtextalpha\xspace}
\let\oldAA\AA
\renewcommand{\AA}{\text{\oldAA}\xspace}
\let\oldtextdegree\textdegree
\renewcommand{\textdegree}{\oldtextdegree\xspace}

\newcommand{\orcid}[2]{\href{http://orcid.org/#2}{#1{\includegraphics[height=10pt]{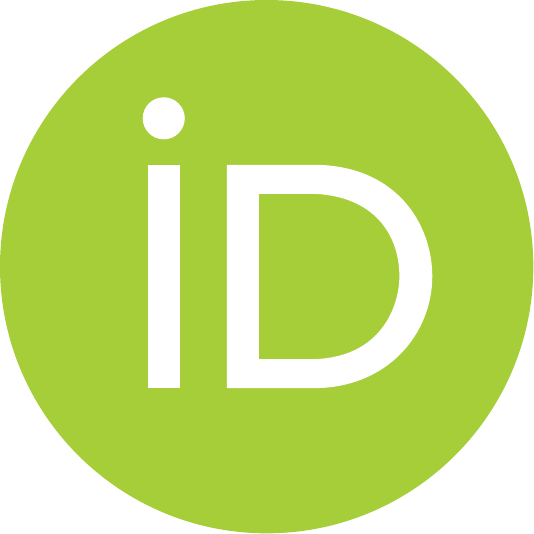}}}}

\newcommand{\kms}{\ensuremath{\mathrm{km\,s^{-1}}}\xspace}
\newcommand{\MSun}{\ensuremath{{\rm M}_\odot}\xspace}
\newcommand{\yr}{\ensuremath{{\rm yr}}\xspace}
\newcommand{\Myr}{\ensuremath{{\rm Myr}}\xspace}
\newcommand{\Gyr}{\ensuremath{{\rm Gyr}}\xspace}
\newcommand{\peryr}{\ensuremath{{\rm yr^{-1}}}\xspace}
\newcommand{\Lsun}{\hbox{\,${\rm L}_\odot$}}
\newcommand{\mum}{\text{\textmu m}\xspace}
\newcommand{\dex}{\text{dex}\xspace}
\newcommand{\kpc}{\text{kpc}\xspace}
\newcommand{\ZH}{\text{[Z/H]}\xspace}
\newcommand{\CO}{\text{[C/O]}\xspace}
\newcommand{\NO}{\text{[N/O]}\xspace}
\newcommand{\FeH}{\text{[Fe/H]}\xspace}
\newcommand{\percm}[1]{\ensuremath{\rm cm^{#1}}\xspace}

\newcommand{\eps}{\ensuremath{\epsilon}\xspace}
\newcommand{\mstar}{\ensuremath{M_\star}\xspace}
\newcommand{\mgas}{\ensuremath{M_\mathrm{gas}}\xspace}
\newcommand{\re}{\ensuremath{R_\mathrm{e}}\xspace}
\newcommand{\NH}{\ensuremath{N_\mathrm{H}}\xspace}
\newcommand{\tauv}{\ensuremath{\tau_\mathrm{V}}\xspace}
\newcommand{\AV}{\ensuremath{A_\mathrm{V}}\xspace}
\newcommand{\xid}{\ensuremath{\xi_\mathrm{d}}\xspace}
\newcommand{\logoh}{\ensuremath{12 + \log\,(\mathrm{O/H})}\xspace}

\newcommand{\nelec}{\ensuremath{n_\mathrm{e}}\xspace}
\newcommandx{\Mout}[2][1=,2=]{\ensuremath{M_{\mathrm{out}{#2}}^{#1}}\xspace}
\newcommandx{\Mdotout}[2][1=,2=]{\ensuremath{\dot{M}_{\mathrm{out}{#2}}^{#1}}\xspace}

\newcommandx{\fluxdcgs}[1][1=-20]{\ensuremath{\mathrm{10^{#1}~erg~s^{-1}~cm^{-2}~\AA^{-1}}}\xspace}
\newcommandx{\fluxcgs}[1][1=-20]{\ensuremath{\mathrm{10^{#1}~erg~s^{-1}~cm^{-2}}}\xspace}
\newcommandx{\powercgs}[1][1=44]{$\times 10^{#1}$~erg~s$^{-1}$\xspace}
\newcommand{\Av}{\ensuremath{A_V}\xspace}

\newcommand{\Te}{\ensuremath{T_\text{e}}\xspace}
\newcommand{\Tiii}{\Te[\ion{O}{iii}]\xspace}
\newcommand{\Tii}{\Te[\ion{O}{ii}]\xspace}
\newcommand{\TSii}{\Te[\ion{S}{ii}]\xspace}
\newcommand{\TSiii}{\Te[\ion{S}{iii}]\xspace}

\newcommand{\Ne}{n$_{\text{e}}$\xspace}

\newcommand{\EWr}{EW$_{\text{0}}$\xspace}

\newcommand{\jwst}{\textit{JWST}\xspace}
\newcommand{\hst}{\textit{HST}\xspace}
\newcommand{\ppxf}{{\sc ppxf}\xspace}
\newcommand{\beagle}{{\sc beagle}\xspace}
\newcommand{\forcepho}{{\sc forcepho}\xspace}
\newcommand{\prospector}{{\sc prospector}\xspace}

\newcommand{\Lyalpha}{\text{Ly\textalpha}\xspace}
\newcommand{\Halpha}{\text{H\textalpha}\xspace}
\newcommand{\Hbeta}{\text{H\textbeta}\xspace}
\newcommand{\Hgamma}{\text{H\textgamma}\xspace}
\newcommand{\Hdelta}{\text{H\textdelta}\xspace}
\newcommand{\Pabeta}{\text{Pa\textbeta}\xspace}
\newcommand{\Hepsilon}{\text{H\textepsilon}\xspace}

\newcommandx{\permittedEL}[6][1=O,2=III,3=,4=,5=,6=]{\text{{#1}\,{\sc {#2}}{#3}{#4}{#5}{#6}}\xspace}
\newcommandx{\semiforbiddenEL}[6][1=O,2=III,3=,4=,5=,6=]{\text{{#1}\,{\sc{#2}}]{#3}{#4}{#5}{#6}}\xspace}
\newcommandx{\forbiddenEL}[6][1=O,2=III,3=,4=,5=,6=]{\text{[{#1}\,{\sc{#2}}]{#3}{#4}{#5}{#6}}\xspace}

\newcommand{\EW}[1]{\text{EW(#1)}\xspace}

\newcommand{\hii}{\permittedEL[H][ii]}

\newcommand{\NV}{\permittedEL[N][v]}
\newcommandx{\NVL}[1][1=1243]{\permittedEL[N][v][\textlambda][#1]}
\newcommandx{\NVall}{\permittedEL[N][v][\textlambda][\textlambda][1239,][1243]}
\newcommand{\NII}{\forbiddenEL[N][ii]}

\newcommand{\NIV}{\semiforbiddenEL[N][iv]}
\newcommandx{\NIVL}[1][1=1486]{\semiforbiddenEL[N][iv][\textlambda][#1]}

\newcommand{\CIV}{\permittedEL[C][iv]}
\newcommandx{\CIVL}[1][1=1550]{\permittedEL[C][iv][\textlambda][#1]}
\newcommand{\CIVall}{\permittedEL[C][iv][\textlambda][\textlambda][1549,][1551]}

\newcommand{\HeII}{\permittedEL[He][ii]}
\newcommandx{\HeIIL}[1][1=1640]{\permittedEL[He][ii][\textlambda][#1]}

\newcommand{\OIII}{\semiforbiddenEL[O][iii]}
\newcommandx{\OIIIL}[1][1=1666]{\semiforbiddenEL[O][iii][\textlambda][#1]}
\newcommand{\OIIIall}{\semiforbiddenEL[O][iii][\textlambda][\textlambda][1661,][1666]}

\newcommand{\OIIIopt}{\forbiddenEL[O][iii]}
\newcommandx{\OIIIoptL}[1][1=5007]{\forbiddenEL[O][iii][\textlambda][#1]}
\newcommand{\OIIIoptall}{\forbiddenEL[O][iii][\textlambda][\textlambda][4959,][5007]}

\newcommand{\NIII}{\semiforbiddenEL[N][iii]}
\newcommandx{\NIIIL}[1][1=1750]{\semiforbiddenEL[N][iii][\textlambda][#1]}
\newcommand{\NIIIall}{\semiforbiddenEL[N][iii][\textlambda][\textlambda][1747--][1754]}

\newcommandx{\CIII}{\semiforbiddenEL[C][iii]}
\newcommandx{\CIIIL}[1][1=1909]{\semiforbiddenEL[C][iii][\textlambda][#1]}
\newcommand{\CIIIall}{\semiforbiddenEL[C][iii][\textlambda][\textlambda][1907,][1909]}

\newcommand{\CIIIp}{\permittedEL[C][iii]}
\newcommand{\NIIIp}{\permittedEL[N][iii]}

\newcommand{\NeIV}{\forbiddenEL[Ne][iv]}
\newcommandx{\NeIVL}[1][1=2424]{\forbiddenEL[Ne][iv][\textlambda][#1]}
\newcommand{\NeIVall}{\forbiddenEL[Ne][iv][\textlambda][\textlambda][2422,][2424]}

\newcommand{\MgII}{\permittedEL[Mg][ii]}
\newcommandx{\MgIIL}[1][1=2803]{\permittedEL[Mg][ii][\textlambda][#1]}
\newcommand{\MgIIall}{\permittedEL[Mg][ii][\textlambda][\textlambda][2796,][2803]}

\newcommand{\NeV}{\forbiddenEL[Ne][v]}
\newcommandx{\NeVL}[1][1=3426]{\forbiddenEL[Ne][v][\textlambda][#1]}
\newcommand{\NeVall}{\forbiddenEL[Ne][v][\textlambda][\textlambda][3346,][3426]}

\newcommand{\OII}{\forbiddenEL[O][ii]}
\newcommandx{\OIIL}[1][1=3727]{\forbiddenEL[O][ii][\textlambda][#1]}
\newcommand{\OIIall}{\forbiddenEL[O][ii][\textlambda][\textlambda][3726,][3729]}

\newcommand{\OIIaur}{\forbiddenEL[O][ii][\textlambda][\textlambda][7320,][7330]}

\newcommand{\SII}{\forbiddenEL[S][ii]}
\newcommandx{\SIIL}[1][1=6725]{\forbiddenEL[S][ii][\textlambda][#1]}
\newcommand{\SIIall}{\semiforbiddenEL[S][ii][\textlambda][\textlambda][6718,][6732]}

\newcommand{\SIII}{\forbiddenEL[S][iii]}
\newcommandx{\SIIIL}[1][1=9068]{\forbiddenEL[S][iii][\textlambda][#1]}

\newcommand{\SIIIall}{\semiforbiddenEL[S][iii][\textlambda][\textlambda][9068,][9532]}

\newcommand{\NeIII}{\forbiddenEL[Ne][iii]}
\newcommandx{\NeIIIL}[1][1=3869]{\forbiddenEL[Ne][iii][\textlambda][#1]}
\newcommand{\NeIIIall}{\forbiddenEL[Ne][iii][\textlambda][\textlambda][3869,][39xx]}

\newcommand{\FeII}{\forbiddenEL[Fe][ii]
[\textlambda][4359]}

\newcommand{\ArIII}{\forbiddenEL[Ar][iii]}
\newcommand{\ArIIIL}{\forbiddenEL[Ar][iii]
[\textlambda][7135]}

\newcommand{\OI}{\forbiddenEL[O][i]}
\newcommand{\OIL}{\forbiddenEL[O][i]
[\textlambda][6300]}

\newcommand{\hda}{\ensuremath{\mathrm{H\text{\textdelta}_A}}\xspace}
\newcommand{\hga}{\ensuremath{\mathrm{H\text{\textgamma}_A}}\xspace}

\usepackage{hyperref}
\hypersetup{
    colorlinks=true,
    linkcolor=blue,
    citecolor=blue,
    filecolor=magenta,      
    urlcolor=cyan,
    }
    
\begin{document}

\title{One cloud is not enough: extreme conditions bias chemical abundances in high-redshift galaxies}
\titlerunning{HOMERUN at high redshift}

\author{
B. Moreschini\inst{\ref{UNIFI},\ref{arcetri},\ref{eso}} \and
F. Belfiore\inst{\ref{arcetri}, \ref{eso}} \and
A. Marconi\inst{\ref{UNIFI},\ref{arcetri}} \and
E. Cataldi\inst{\ref{UNIFI}, \ref{arcetri}} \and
M. Curti\inst{\ref{eso}} \and
A.
Amiri\inst{\ref{IPM}, \ref{UOA}} 
\and
A. Feltre\inst{\ref{arcetri}} \and
F. Mannucci\inst{\ref{arcetri}} \and
E. Bertola\inst{\ref{arcetri}}\and
C. Bracci\inst{\ref{UNIFI},\ref{arcetri}, \ref{eso}}\and
M. Ceci\inst{\ref{UNIFI},\ref{arcetri}}\and
A. Chakraborty\inst{\ref{arcetri},\ref{IIA}}\and
G. Cresci\inst{\ref{arcetri}} \and
Q. D'Amato\inst{\ref{arcetri}} \and
E. Di Teodoro\inst{\ref{UNIFI},\ref{arcetri}}\and
M. Ginolfi\inst{\ref{UNIFI},\ref{arcetri}} \and
I. Lamperti\inst{\ref{UNIFI},\ref{arcetri}} \and
C. Marconcini\inst{\ref{UNIFI},\ref{arcetri}}\and
M. Scialpi\inst{\ref{trento},\ref{arcetri},\ref{UNIFI}}\and
L. Ulivi\inst{\ref{trento},\ref{arcetri},\ref{UNIFI}}\and
M. V. Zanchettin\inst{\ref{arcetri}}
}

\institute{
\label{UNIFI} Università di Firenze, Dipartimento di Fisica e Astronomia, via G. Sansone 1, 50019 Sesto Fiorentino, Florence, Italy \and
\label{arcetri} INAF -- Arcetri Astrophysical Observatory, Largo E. Fermi 5, I-50125, Florence, Italy \and
\label{eso} European Southern Observatory, Karl-Schwarzschild Straße 2, D-85748 Garching bei München, Germany \and
\label{trento} University of Trento, Via Sommarive 14, I-38123 Trento, Italy
\and
\label{IPM}
School of Astronomy, Institute for research in fundamental sciences (IPM), Tehran, P.O. Box 19395-5531, Iran
\and
\label{UOA}
Department of Physics, University of Arkansas, 226 Physics Building, 825 West Dickson Street, Fayetteville, AR 72701, USA
\and
\label{IIA} 
Indian Institute Of Astrophysics, 100 Feet Rd, Santhosapuram, 2nd Block, Koramangala, Bengaluru, Karnataka 560034, India
}

\authorrunning{B. Moreschini et al.}
\date{}
\abstract{
Since its launch, JWST has opened an unprecedented opportunity to characterise the ionised interstellar medium (ISM) of high-redshift galaxies using well-established rest-frame ultra-violet (UV)/optical diagnostics from the local Universe. At the same time, these observations challenge the validity of such classical methods when applied to the extreme and diverse environments typical at high redshift. In this paper, we present an in-depth analysis of the ISM in three representative case studies at $z=2 - 6$ (MARTA 4327, the Sunburst Arc and RXCJ2248-ID) conducted within a multi-cloud photoionisation modelling framework (HOMERUN), which models integrated galaxy spectra as linear combinations of photoionised regions spanning a wide range of ionisation and density conditions. We show that even a small fraction of unresolved high-density ($n_\mathrm{e}=10^4 - 10^7 \ \mathrm{cm^{-3}}$) clumps can contribute more than half of the observed flux of auroral lines such as [O III] $\lambda$4363 and [O II] $\lambda\lambda$7323,7332, while only negligibly to standard optical density tracers. As a result, $T_{\mathrm{e}}$-method metallicities can be underestimated by $\sim 0.15 - 0.3$ dex, as for MARTA 4327. By modelling rest-frame UV and optical data simultaneously, we demonstrate that discrepancies between abundance estimates obtained from diagnostics tracing different zones do not necessarily imply chemical inhomogeneities. In RXCJ2248-ID, the disagreement between UV and optical N/O may naturally arise from ionisation and density structure alone. In contrast, we find evidence for genuine chemical stratification in the Sunburst Arc, where a matter-bounded component enriched in nitrogen ($\log(\mathrm{N/O})=-0.46$) coexists with a chemically normal ($\log(\mathrm{N/O})= - 1.39$), ionisation-bounded one. Finally, we argue that high-ionisation lines such as He II $\lambda$1640 and C IV $\lambda\lambda$1548,1551 can be explained within a pure star-formation scenario invoking matter-bounded regions, consistent with a non-negligible Lyman-continuum leakage. However, in the case of RXCJ2248-ID, we cannot rule out a minor contribution from an active galactic nucleus based solely on the observed emission-line fluxes. Together, these results indicate that classical diagnostics can be significantly biased in high-redshift galaxies and that self-consistent, physically motivated tools are therefore essential to properly interpret the complex ISM conditions and chemical enrichment in the early Universe.
}

   \keywords{galaxies: high-redshift – galaxies: evolution – galaxies: abundances}

   \maketitle
%
\section{Introduction}
\label{sec:introduction}
The interstellar medium (ISM) plays a major role in galaxy evolution: it is where stars form, and the stage for all feedback and regulatory processes governing star formation and interactions with the environment \citep{Kennicutt1998, Schinnerer2024}. 
A large body of observations from the last decade has demonstrated that the high-$z$ ISM differs from its local counterpart, showing higher ionisation conditions, harder ionising spectra, lower gas-phase metallicities, and higher densities \citep[e.g.,][]{Shapley2015,Sanders2016,Steidel2016, Strom2017, Maseda2017, Sanders2021, Mainali2023, Sanders2023}. However, due to observational limits, rest-frame optical spectroscopy was confined mostly to Cosmic Noon.
Thanks to its sensitivity and wavelength coverage, JWST has revolutionised our view of the high-$z$ ISM by enabling detailed characterisation of galaxies from Cosmic Noon to Cosmic Dawn \citep[e.g.,][for recent reviews]{Ellis2025, Matthee2025, Stark2025}. Such detailed spectroscopic information has highlighted key differences in gas density, ionising spectra, and chemical abundance patterns.

Firstly, JWST has revealed a population of luminous, compact galaxies at $z\sim10$, characterised by low metallicities, high star-formation rates, and possibly powered by intense starburst episodes \citep[e.g.,][]{Bunker2023, Castellano2024, Curti2025a, Napolitano2025, Naidu2025, Scholtz2025}. Their spectra often show very high-ionisation species like N IV] $\lambda\lambda1483,1486$, C IV $\lambda\lambda$1548,1551, and He II $\lambda$1640, which need photons with energies above 47.5, 47.8, and 54.4 eV, respectively. These extreme emission features are difficult to reproduce with standard stellar population models, and often require invoking strong bursts of young, metal-poor stars or alternative hard sources, such as active galactic nuclei (AGN), X-ray binaries, Wolf–Rayet (WR) stars \citep[e.g.,][]{Shirazi2012, Senchyna2017, Nanayakkara2019, Schaerer2019, Tang2023}, or even super- and very-massive stars \citep[SMS, VMS, with $M\sim10^3-10^5 \ {\rm M}_{\odot}$ and $\sim10^2-10^3 \ \mathrm{M}_{\odot}$;][]{Charbonnel2023, Vink2023}.

The growing number of detections of AGN-like features (such as high-ionisation lines or broad components) at high redshift raises the issue of disentangling the contributions of AGN activity from that of metal-poor stellar sources. The classical method for separating AGN from stellar photoionisation relies on rest-frame optical diagrams, like the Baldwin-Phillips-Terlevich (BPT) diagram \citep[][]{Baldwin1981, Kewley2006}. However, the evolving ISM conditions at high redshift cast doubt on its applicability \citep[e.g.,][]{Kewley2013,Hirschmann2017, Hirschmann2019, Cleri2025, Scholtz2025c}. Moreover, classical methods often struggle to distinguish mixed contributions from AGN and star formation (with some exceptions, e.g., \citealt{DAgostino2019} for integral field data), and to robustly constrain ISM conditions in such cases. Recent spectral fitting tools like BEAGLE-AGN \citep{VidalGarcia2024} and \texttt{Cue} \citep{Li2025a} are however making progress in this direction. 

Among high-$z$ galaxies with unusual properties, N-rich systems have attracted particular attention. These galaxies have elevated nitrogen abundances ($\sim0.5-1.5$ dex) as inferred from their rest-frame ultra-violet (UV) lines \citep[e.g.,][for compilations]{Ji2025a, Morel2025}. 
One interpretation links these galaxies to the progenitors of local globular clusters, given their similar abundance patterns \citep[e.g.,][]{Charbonnel2023, Isobe2023a, Senchyna2024, Ji2025a}. Physically, this may be connected to the presence of massive stars, whose stellar feedback (through -also temporally- differential winds and consequent shocks) could enhance observed N/O abundances in localised, chemically enriched regions \citep[e.g.,][]{Flury2025,Rizzuti2025}. Some studies suggest additional contributions from multiple bursts of star formation or tidal disruption events \citep[e.g.,][]{Cameron2023b, Kobayashi2024, Topping2024, Watanabe2024, Bhattacharya2025, McClymont2025}. Furthermore, in dense stellar clusters, stellar collisions may promote the formation of black holes, potentially seeding intermediate-mass black holes \citep[e.g.,][]{Rantala2025}. Such a connection between black holes and N-emitters provides a possible explanation for the AGN signatures observed in many of these systems \citep{Isobe2025a}.

Finally, theoretical predictions, simulations, pre-JWST and JWST-based studies found evidence for increasing electron density with redshift, using both UV and optical diagnostics \citep[e.g.,][]{vanderWel2014, Sanders2016, Maseda2017, Isobe2023, Abdurro2024, Topping2025, Martinez2025}. In particular, recent works found densities sometimes $\gtrsim10^5$ cm$^{-3}$. Most common rest-frame optical density diagnostics involve low critical-density transitions (like [O II] $\lambda\lambda$3727,3729 or [S II] $\lambda\lambda$6717,6731), which are not sensitive to such high-density clumps. This aspect becomes even more crucial in high-redshift environments, where the ISM may exhibit strong density stratification ($n_{\rm e}\sim10-10^5$ cm$^{-3}$, e.g., \citealt{Harikane2025a, Usui2025}; also, e.g., \citealt{Mingozzi2022} for a high-redshift-analogues perspective).
For instance, \citet{Hayes2025} suggested that anomalous nitrogen enhancement might partially result from incorrect density assumptions: if densities are underestimated, collisional boosting of auroral-to-nebular oxygen ratios may mimic low metallicity, inflating N/O. \citet{Martinez2025} further explored this idea, conducting a detailed analysis on a sample of local and high-$z$ galaxies with optical and far UV diagnostics. Their analysis confirmed that classical metallicity estimates can be severely biased, and high-N/O galaxies may not be as enhanced as previously thought. Conversely, using a two-zone far-IR-optical model combining ALMA and JWST data, \citet{Harikane2025} showed that neglecting contributions from very low-density (below [S II] sensitivity) gas can also lead to metallicities being underestimated by $\simeq0.4$ dex in $z\simeq8$ galaxies.

In order to improve our understanding of the sources of stellar feedback and how star formation and AGN activity proceed in the early Universe, it is therefore critical to simultaneously constrain ISM conditions and model the ionising source by fitting all available spectral data over a wide wavelength range. In this paper, we propose a unified framework for modelling the ionised ISM at high redshift by using the multi-cloud photoionisation framework called HOMERUN (Highly Optimised Emission-line Ratios Using photoioNisation; \citealt{Marconi2024}). In this approach, a linear combination of photoionisation models is used, with minimal assumptions, to fit the emission lines originating from a variety of ISM conditions within spatially integrated galaxy spectra.

To demonstrate the power and flexibility of the method, we present here an in-depth analysis of the physical and chemical properties in three well-studied high-redshift galaxies: MARTA 4327 ($z= 2.2$, \citealt{Cataldi2025, Curti2025}), the Sunburst Arc ($z=2.3$, \citealt{Pascale2023, RiveraThorsen2024, Welch2025}), and RXCJ2248-ID ($z=6.1$, \citealt{Topping2024, Berg2025}). These objects were selected for their high-S/N observations, data available over a wide wavelength coverage (optical to near-IR for MARTA 4327, UV/optical for the other two) and diversity in metallicity ($12 + \log($O/H$)\sim8.2, \ 7.9, \ 7.7$), stellar mass ($\log(M/\mathrm{M}_{\odot})\sim  9, \ 7 \ , \ 8$), and electron densities in the range log($n_{\mathrm{e}}$/cm$^{-3})\sim2-5$ \citep{Vanzella2020, Welch2025, CrespoGo2025, Cataldi2025}. A common feature is that they all host WR stars \citep{RiveraThorsen2024, Curti2025, Berg2025}, allowing us to probe the environment of massive stellar populations over a wide range of ISM conditions.
In addition, both the Sunburst Arc and RXCJ2248-ID are known to be over-enriched in N/O. The former also shows confirmed chemical stratification \citep{Pascale2023}, and the latter is a strong C IV and N IV emitter \citep{Topping2024}.

This paper is structured as follows: in Sec. \ref{sec:homerun}, we describe our modelling assumption and $T_{\mathrm{e}}$ abundance computation. Sections \ref{sec:marta section}, \ref{Sunburst models}, and \ref{sec:Topping_models} present our application of HOMERUN to each source. In Sec. \ref{sec:discussion}, we discuss the impact of ISM inhomogeneities on abundance determinations and how to interpret the hard ionisation field observed within high-$z$ sources. Our conclusions are summarised in Sec. \ref{sec:conclusion}.

\section{Methods: multi-cloud photoionisation modelling}
\label{sec:homerun}

HOMERUN is a photoionisation modelling framework developed to capture the complexity of the ISM through a multi-cloud approach. 
In particular, observed emission lines are modelled with a linear combination of constant-density CLOUDY \citep{1983hbic.book.....F} models (hereafter `single-cloud' models), each characterised by a specific ionisation parameter and gas density, but sharing the same ionising spectrum and chemical composition.
The key novelty of HOMERUN lies in considering the weights of this combination as free parameters, directly inferred by fitting the observed emission-line fluxes.
In previous works, HOMERUN was applied to model a different range of objects from local HII regions and AGN to high-redshift sources \citep{Marconi2024,  Castellano2025,Ceci2025, Marconcini2025,Tripodi2025}, highlighting the need for self-consistent, multi-line photoionisation modelling to properly
capture the complex emission in such objects.
Below, we briefly summarise the model assumptions and methodology, and refer the reader to \citet{Marconi2024} for further details.

\subsection{Grids of CLOUDY models}\label{subsec:grids of cloudy models}

The single-cloud models that constitute the library used in this work were computed with version 23.01 of CLOUDY \citep{Gunasekera2023}.  
For our analysis, we use both HII region and narrow-line AGN models. The model parameters are listed in Appendix \ref{app:AppA}, and in the rest of this section we give a brief overview of some of the model assumptions.


For the HII region models, we used as input spectra the BPASS simple stellar population models \citep{Stanway2018}, including binaries, assuming a \citet{Kroupa2002} initial mass function extending up to 300 M$_\odot$, and a single burst of star formation.
We considered a range of ages within $\log(t\mathrm{_{age}/yr}) = [6, 8]$, and stellar metallicities $\log(Z_\star) = -4, -2.7, -1.7$. 
While we do not fix the value of gas-phase metallicity to the stellar one, we exclude extreme offsets as reported in Appendix \ref{app:AppA}.
Our narrow-line AGN models have been presented in \citet{Ceci2025}. The shape of their ionising radiation is defined by a fixed UV slope $\alpha_{\rm UV}=-0.5$ and X-ray component with slope $\alpha_{\rm X}=-1$. We varied the X-ray to UV ratio $\alpha_{\rm ox}$ and the exponential cut-off at temperature $T_{\rm max}$ (see Appendix \ref{app:AppA}). 
We adopted the solar abundance pattern from \citet{Asplund2021}, and scaled the N and C abundances with metallicity following the relations by \citet{Nicholls2017}, with an additional offset of $+0.2$ dex in case of nitrogen. We include both models with and without depletion onto dust grains, assuming CLOUDY’s default values, and dust-grain physics. 
All models were computed assuming a plane-parallel and open geometry, and have constant density.

In addition, we computed HII region matter-bounded models, defined by stopping the computation when the medium starts becoming optically thick to He II (specifically when the optical depth at 4.0 Ryd reaches 0.5). Such a criterion was chosen to maximise the He II emission and therefore facilitate the fitting of sources where He II lines are unusually strong. All these models have an escape fraction of Lyman-continuum (LyC) photons of near unity (except for a few models with $\log U$<-3). 

\subsection{Multi-cloud fitting with HOMERUN}

HOMERUN builds the best-fit model with a single ionising continuum and abundance pattern. For a given ionising continuum and gas-phase metallicity, we construct a so-called `multi-cloud model' from a linear combination of single clouds of varying gas density and ionisation parameter $U$ (defined as the ratio of hydrogen-ionising photon flux and the hydrogen density). The optimal combination is found via non-negative least squares minimising a $\chi^2$-like loss function, hereafter labelled as $\mathcal{L}$.
This procedure is then iterated across all metallicities and continua, and the best-fit solution corresponds to the global minimum of $\mathcal{L}$. 

As error term $\sigma$ in the calculation of $\mathcal{L}$  we adopt the maximum between the observational error and 10\% of the observed line flux. This minimum error term accounts for possible residual systematics in the modelling which are present even at high signal-to-noise. Uncertainties on model-derived quantities are computed considering all solutions with $\mathcal{L} \leq 1.25 \, \mathcal{L}_{\mathrm{min}}$. 
While this threshold is arbitrary and chosen to be conservative, it does not affect the best-fit values, identified by the model with $\mathcal{L_\mathrm{min}}$.

For the Sunburst Arc and RXCJ2248-ID, we also discuss models constituted by a linear combination of different continua and/or abundance patterns (Sec. \ref{sec: Sunburst fiducial model}, Sec. \ref{sec: Topping multi comp mods}). We refer to these as `multi-component' models. Each such model is individually multi-cloud, including a range of densities and ionisation parameters.


To account for non-solar abundance patterns, we assume that line fluxes of metal ions scale linearly with abundance within the range considered. We allowed the emission lines of helium and all metals but oxygen to be rescaled by $\pm$ 1 dex with respect to solar, treating the scaling as an additional free parameters during the fit. For C and N we consider variations with respect to the \cite{Nicholls2017} relation rather than the solar value. In Appendix \ref{app:AppA} we discuss the impact of our assumptions in inferring chemical abundances.

During the fit we also infer the V-band dust attenuation $A_{\mathrm{V}}$, adopting a given attenuation curve. We estimate the total attenuation from all observed lines, without relying solely on Balmer lines or assuming fixed temperature and density values.
Recent studies highlighted that attenuation curves in high-redshift galaxies may differ significantly from those in the local Universe \citep[e.g.,][]{Markov2025, Sanders2025a, Reddy2025}. 
However, we verified that our main conclusions remain robust when adopting different locally calibrated prescriptions \citep{Cardelli1989, Calzetti2000, Gordon2003}. Finally, after each fit, we checked post-facto that the best-fit model did not predict any emission lines which should have otherwise been observed in the data given their predicted flux.

\subsection{Comparison with the $T_{\mathrm{e}}$-method} \label{Te-method sec}

We benchmark the abundances inferred from the HOMERUN multi-cloud models with the $T_{\mathrm{e}}$-based abundances, calculated under standard assumptions. In particular, we use \texttt{PyNeb} \citep{Luridiana2015}, adopting the same atomic data used in the CLOUDY photoionisation models (Tab. \ref{tab:atomic_parameters}).
As a first step, we simultaneously determine the electron density and the temperature $T$(O II) of the low-ionisation zone using \texttt{getCrossTemDen}, based on [S II] $\lambda\lambda$6717,6731 and the [O II] $\lambda\lambda$3727,3729 / [O II] $\lambda\lambda$7323,7332 ratios. We then estimate the high-ionisation temperature $T$(O III) from [O III] $\lambda$4363 / [O III] $\lambda$5007 using \texttt{getTemDen} and the previously determined density. Given the high critical densities of the [O III] lines involved in this step, changing the assumed value for the gas density in this second step does not impact the derived abundances for typical sources, except for RXCJ2248-ID (see Sec. \ref{UV diagnostic section}). We then measure the total oxygen abundance as the sum of the O$^{+}$ and O$^{++}$ abundances obtained with \texttt{getIonAbundance}.

For abundances of other elements, we follow similar standard prescriptions. 
To account for unseen ionisation states, when possible we adopt the ionisation correction factors (ICFs) following \citet{ArellanoCo2024}, based on an analysis of local high-redshift analogues from the CLASSY survey. Details on the abundance derivation procedure are provided in Appendix \ref{app:AppA}.
Uncertainties are estimated by repeating the calculation 200 times, perturbing the line fluxes with random values drawn from a normal distribution with a standard deviation equal to the observed errors. 


\section{The importance of density fluctuations: M~4327}
\label{sec:marta section}
\begin{figure*}[ht]
    \includegraphics[width=\linewidth]{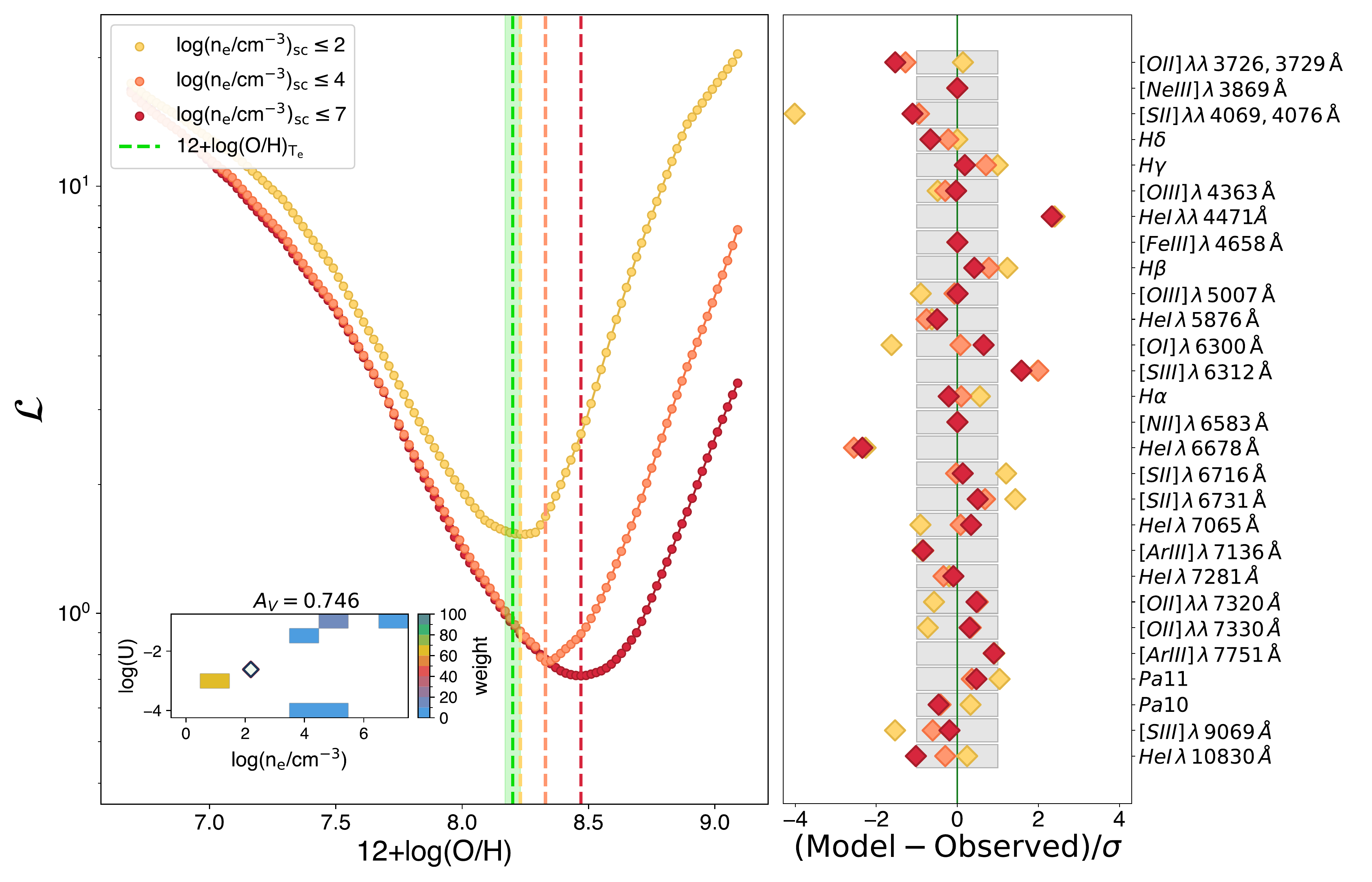}
    \caption{\textit{Left}: Variation of the loss function $\mathcal{L}$ with model metallicity for HOMERUN models of M~4327 considering different maximum densities of the single-cloud models adopted in the fit. The yellow (red) curve corresponds to the lowest (highest) density limits. Yellow, orange and red dashed lines mark the minimum of $\mathcal{L}$, while the green dashed line represents the metallicity obtained with the $T_{\mathrm{e}}$-method. The inset shows the grid of single-cloud models which contribute to the best fit in the highest-density case coloured by their weight, with the white diamond corresponding to the weighted (harmonic) average. On the top of the inset we report the best-fit value for $A_{\mathrm{V}}$, assuming the attenuation law by \citet{Cardelli1989}. \textit{Right}: Comparison between our fiducial best-fit model predictions and observed fluxes for all lines included in the fit, and the three models discussed in Sec. \ref{sec:marta_HR} (coloured diamonds). Grey bars show $1\sigma$ deviation, where $\sigma$ is the maximum between the relative error and 10\% of the observed flux, assumed as the minimum acceptable discrepancy.}
    \label{M4327 chisq}
\end{figure*}

As a case-study providing valuable insights into the impact of density inhomogeneities on abundance determinations at high redshift, we present the fit for MARTA 4327 (hereafter M~4327). M~4327 is a star-forming galaxy ($\log(M_{\star}/\mathrm{M}_{\odot})=9.36$ $\mathrm{SFR}=7.28 \ \mathrm{M_{\odot}yr^{-1}}$), 
at $z = 2.2$ hosting a population of WR stars \citep{Curti2025} part of the MARTA (Measuring Abundances at high Redshift with the $T_{\mathrm{e}}$ Approach) programme (PID 1879, PI: Curti). MARTA is an extremely deep JWST/NIRSpec survey which targeted galaxies at Cosmic Noon \citep{Cataldi2025, Curti2025}.
M~4327 shows high signal-to-noise detections of multiple auroral lines ([S II] $\lambda\lambda$4068,4076, [O III] $\lambda$4363, [S III] $\lambda$6312, [O II] $\lambda\lambda$7323,7332), hydrogen and helium recombination lines, the density-sensitive He I $\lambda$10831 multiplet
\citep[e.g.,][]{Izotov2014, Aver2015, Berg2025} and [S II] $\lambda\lambda$6717,6731 doublet.
The combination of NIR-to-optical coverage and deep spectroscopy enables a robust comparison between model-based properties and those derived through classical methods when UV transitions are not available.
A dedicated study of this object is presented in \citet{Curti2025}, and we adopt their line fluxes here.

\subsection{A multi-cloud HOMERUN model with different density thresholds}
\label{sec:marta_HR}

We fitted with HOMERUN all available lines except the Balmer lines of order higher than H$\delta$ and O I $\lambda$8446
. The higher-order Balmer lines are either blended with other features or too faint and highly sensitive to uncertainties in stellar continuum subtraction.
O I $\lambda$8446, on the other hand, is influenced by complex processes such as dielectronic recombination, stellar fluorescence, and Ly$\beta$ continuum pumping (see, e.g., \citealt{Choe2025} and \citealt{Curti2025} for a comprehensive discussion). 
We therefore decided to exclude O I $\lambda$8446 from our fits.
For lines with ratios fixed by atomic physics (e.g., [O III] $\lambda\lambda$4959,5007), we only included the brightest component. $A_{\mathrm{V}}$ was left free during the fit, adopting the \citet{Cardelli1989} attenuation curve for consistency with \citet{Curti2025}.
\begin{table*}
\caption{Derived ISM properties for M~4327 both via the best-fit models and $T_{\mathrm{e}}$-method}
\centering
\renewcommand{\arraystretch}{1.3} 

\begin{tabular}{lccccccc}
\toprule
\midrule
& \multicolumn{6}{c}{HOMERUN models} & $\mathrm{T_{e}}$-method \\
& \multicolumn{2}{c}{$n_\mathrm{e,max}=10^2$ cm$^{-3}$} 
& \multicolumn{2}{c}{$n_\mathrm{e,max}=10^4$ cm$^{-3}$} 
& \multicolumn{2}{c}{$n_\mathrm{e,max}=10^7$ cm$^{-3}$} 
& \\
\midrule
$<\log U$>  & & $-2.25^{+0.07}_{-0.19}$ & & $-2.47^{+0.06}_{-0.14}$ & & $-2.62^{+0.15}_{-0.03}$ & $-2.572\pm0.003^{a}$ \\
$<\log(n_{\mathrm{e}}/\mathrm{cm}^{-3})>$  & & $2.0^{+0.0}_{-0.4}$ & & $1.9^{+0.8}_{-0.6}$ & & $2.2^{+0.8}_{-0.9}$ & $1.9^{+0.2 \ b}_{-0.4}$\\
$12+\log(\mathrm{O}/\mathrm{H})$  & & $8.23^{+0.10}_{-0.13}$ & & $8.33^{+0.16}_{-0.09}$ & & $8.47^{+0.11}_{-0.15}$ & $8.20\pm0.03$ \\
$\log(\mathrm{Ne}/\mathrm{O})$ & & $-0.53^{+0.05}_{-0.05}$ & & $-0.56^{+0.02}_{-0.06}$ & & $-0.598^{+0.072}_{-0.015}$ & $-0.561\pm0.014$\\
$\log(\mathrm{N}/\mathrm{O})$  & & $-1.37^{+0.09}_{-0.05}$ & & $-1.36^{+0.04}_{-0.06}$ & & $-1.40^{+0.07}_{-0.07}$ & $-1.17\pm0.02$\\
$\log(\mathrm{S}/\mathrm{O})$   & & $-1.74^{+0.06}_{-0.06}$ & & $-1.72^{+0.02}_{-0.09}$ & & $-1.80^{+0.13}_{-0.06}$ & $-1.76\pm0.03^{c}$\\
$\log(\mathrm{Ar}/\mathrm{O})$  & & $-2.46^{+0.09}_{-0.09}$ & & $-2.49^{+0.03}_{-0.10}$ & & $-2.57^{+0.14}_{-0.08}$ & $-2.49\pm0.03$\\
$\log(\mathrm{Fe}/\mathrm{O})$  & & $-2.270^{+0.072}_{-0.016}$ & & $-2.16^{+0.04}_{-0.09}$ & & $-2.19^{+0.06}_{-0.09}$ & $-2.07^{+0.14}_{-0.15}$\\
$12+\log(\mathrm{He}/\mathrm{H})$  & & $10.869^{+0.039}_{-0.010}$ & & $10.87^{+0.06}_{-0.03}$ & & $10.89^{+0.09}_{-0.02}$ & $10.87\pm 0.02^{d}$\\
$A_\mathrm{V}^{e}$  & & $0.782^{+0.030}_{-0.006}$ & & $0.724^{+0.091}_{-0.014}$ & & $0.75^{+0.08}_{-0.02}$ & $0.75^f$ \\
\midrule
\bottomrule
\end{tabular}
\tablefoot{
\tablefoottext{a}{Derived from [O III] $\lambda$5007/[O II] $\lambda\lambda$3727,3729 using the calibration by \citet{Berg2019}.} 
\tablefoottext{b}{Density from [S II] doublet, as described in Sec. \ref{Te-method sec}.} 
\tablefoottext{c}{Assuming $T$(O III) and $T$(O II), as described in Appendix \ref{app:AppA}. If we instead assumed  $T$(S III) and $T$(S II) for the high- and low-ionisation zones, we would obtain $\log(\mathrm{S}/\mathrm{O})=-1.5\pm 0.2$} (Appendix \ref{app:AppA}).
\tablefoottext{d}{Estimated by \citet{Curti2025} with a MCMC procedure neglecting the contribution of doubly ionised helium.} 
\tablefoottext{e}{Assuming \citet{Cardelli1989} for consistency with \citet{Curti2025}.} 
\tablefoottext{f}{To best compare model properties with those derived with standard methods from line ratios, we adopted the HOMERUN best-fit A$_V$ for our fiducial model, i.e. the highest density case.}
}
\label{tab:Marta properties}
\end{table*}

\begin{figure}[h]
    \includegraphics[width=\linewidth]{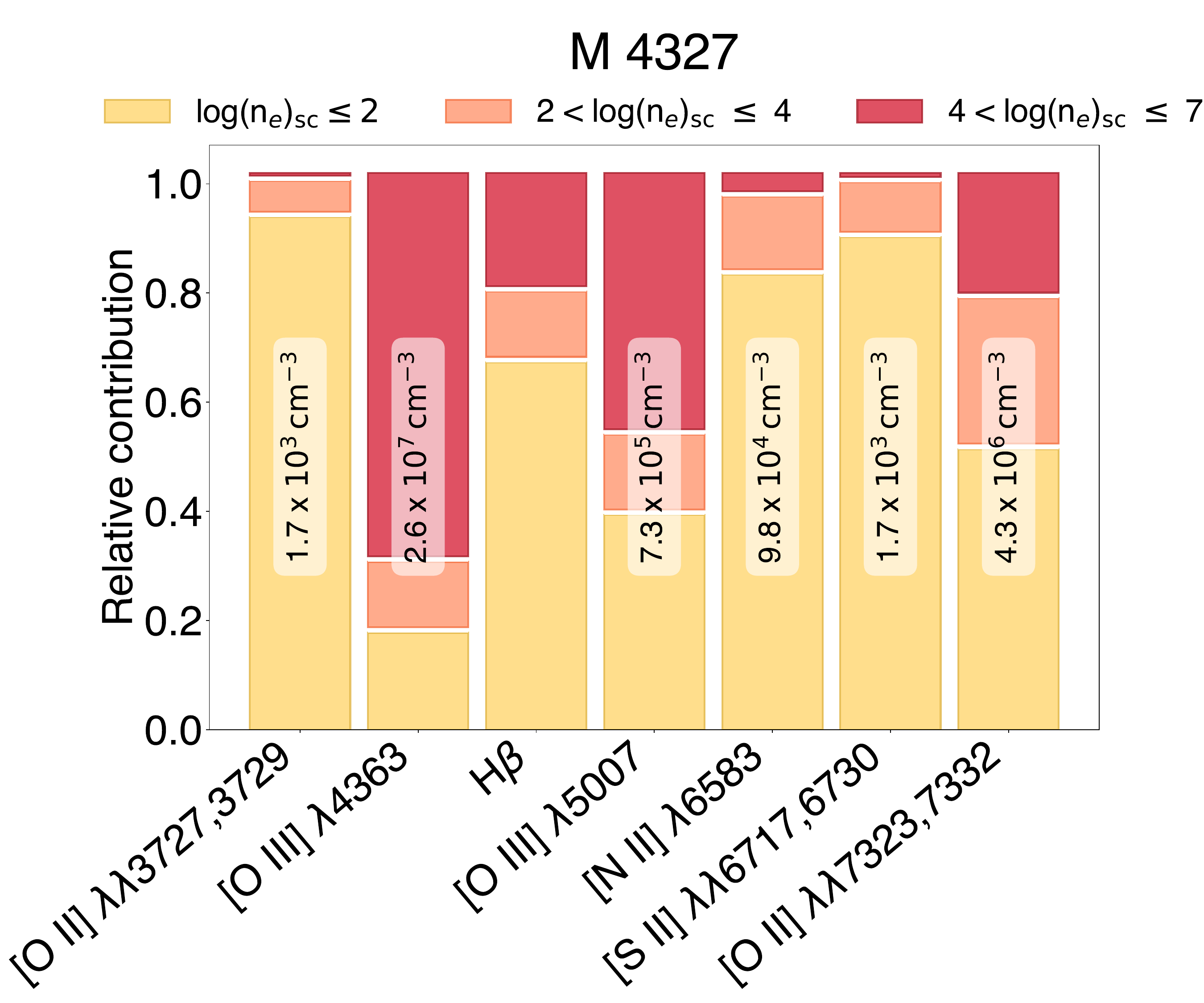}
    \caption{Relative contribution of single-cloud models with density $n_\mathrm{e}\leq10^2$ cm$^{-3}$ (yellow), $10^2 \ \mathrm{cm^{-3}}<n_e\leq10^4$ cm$^{-3}$ (orange), and $n_\mathrm{e}\leq10^7$ cm$^{-3}$ (red), to the main emission lines (predicted by the best-fit model) involved in the determination of ISM chemical and physical properties in M~4327. Within each bar, the box reports the critical density for the corresponding transition, computed with \texttt{PyNeb} assuming T$=12000$ K and $n_\mathrm{e}=200$ cm$^{-3}$. In the case of doublets, only the lowest critical density is shown. Highest-density models contribute only $\simeq20\%$ of the H$\beta$ flux.}
    \label{MARTA 4327 rel contr}
\end{figure} 

To investigate the effect of density inhomogeneities, we performed three HOMERUN fits exploring different ranges of hydrogen densities. In particular, we include a range of models spanning density from 1 $\rm cm^{-3}$ to a varying maximum value:
\begin{enumerate}
    \item A model allowing very high-density substructure, with densities varying up to $\log(n_{\mathrm{e}}/\mathrm{cm}^{-3}) = 7$. This is our fiducial model;
    \item An intermediate model restricting the grid to $\log(n_{\mathrm{e}}/\mathrm{cm}^{-3}) \leq 4$, corresponding to the maximum density probed by the [S II] $\lambda\lambda$6716,6731 doublet;
    \item A low-density model with $\log(n_{\mathrm{e}}/\mathrm{cm}^{-3}) \leq 2$, approximately matching the density measured from [S II] (Table \ref{tab:Marta properties}).
\end{enumerate}

The left panel of Figure \ref{M4327 chisq} shows the variation of the loss function $\mathcal{L}$ as a function of model gas-phase metallicity for the different density cuts.
Comparing these results with the metallicity derived via the $T_{\mathrm{e}}$-method (Sec. \ref{sec:marta_impact}), we found that excluding high-density models led to a best-fit metallicity consistent with the $T_{\mathrm{e}}$ estimate.
In contrast, the inclusion of high-density single-cloud models increased the best-fit metallicity by up to $0.24$ dex, but significantly improved the fit quality.
In fact, as shown in the right panel of Figure \ref{M4327 chisq}, the fiducial model is able to simultaneously reproduce most considered emission lines within 10\% of their observed values or the relative error, with the exception of 
He~I $\lambda$4471
and He I $\lambda$6678.
In addition, our fiducial-model prediction for O I $\lambda$8446 is consistent with the observed value, although we do not fit this line.
\begin{figure*}[ht]
    \centering
    \includegraphics[width=0.45\textwidth]{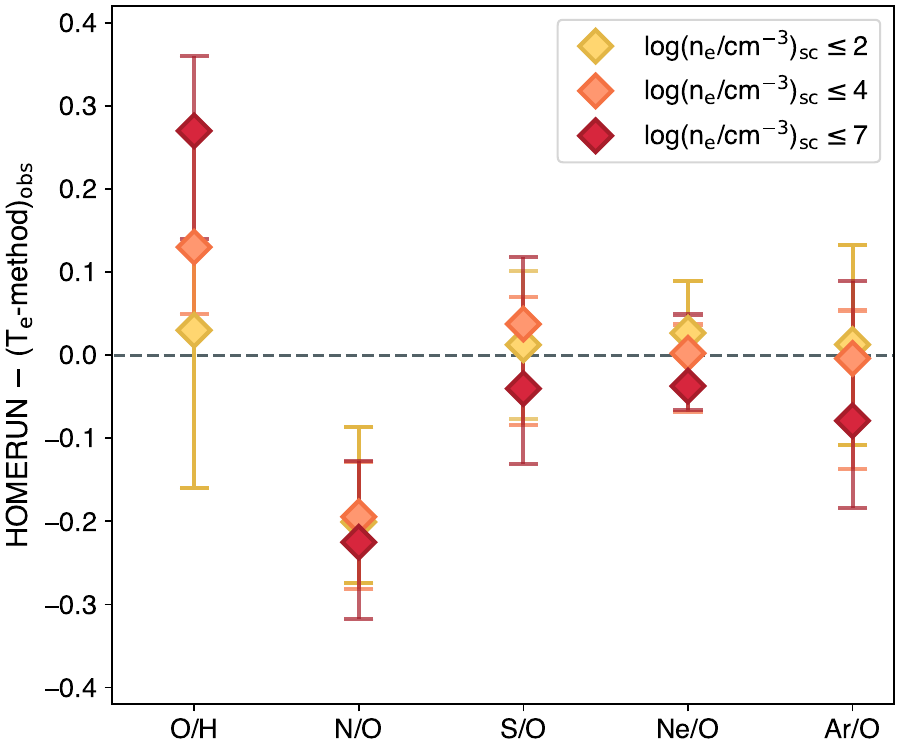}
    \hspace{0.05\textwidth}
    \includegraphics[width=0.45\textwidth]{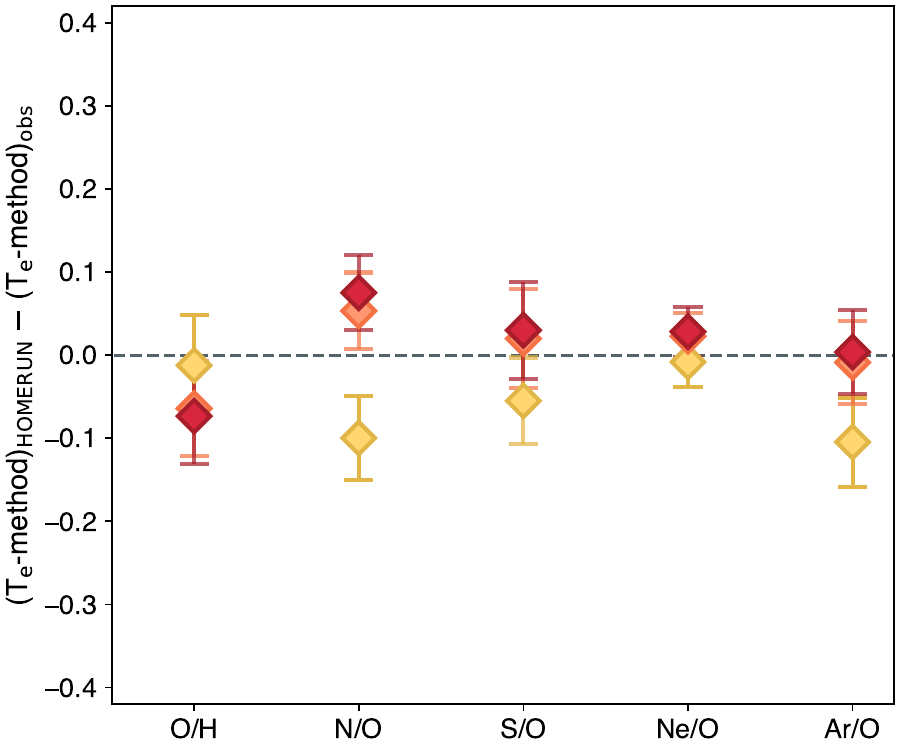}
    \caption{\textit{Left}: discrepancy between model and $T_{\mathrm{e}}$ elemental abundances for M~4327. The yellow, orange and red diamonds correspond to increasing the maximum density allowed for single-cloud models during the fit. \textit{Right}: same, but comparing $T_{\mathrm{e}}$-abundances from observed line fluxes and those from the fluxes predicted by the best-fit model.}
    \label{Abund M4327}
\end{figure*}

The inset panel in Figure \ref{M4327 chisq} illustrates the single-cloud models contributing to the best fit in the highest-density case, their respective weights normalised to H$\beta$, and the weighted (harmonic) average of ionisation parameter and density (shown with the open diamond).
Over all acceptable configurations, the number of single-cloud models contributing to the best-fit fiducial model remains roughly constant, ranging from 6 to 8 in $\simeq80\%$ of the cases.
We find that, even though single-cloud models with $\log(n_{\mathrm{e}}/\mathrm{cm^{-3}})>4$ are responsible for only $\simeq20$\% of the total flux of H$\beta$, they are important in reproducing specific emission lines involved in metallicity determination.

To quantify the effect of density fluctuations, we report in Figure \ref{MARTA 4327 rel contr} the relative contribution of high- and low-density single-cloud models to the main emission lines used to derive oxygen and nitrogen abundances, temperatures and density for the best-fit model of M~4327.
While low critical density or low-ionisation species like [O II] $\lambda\lambda$3727,3729, [S II] $\lambda\lambda$6717,6731 and [N II] $\lambda$6583 are mainly emitted by low-density gas, models with $\log(n_{\mathrm{e}}/\mathrm{cm^{-3}})>2$ are responsible for up to $\simeq80\%$ and $\simeq50\%$ of [O III] $\lambda$4363 and [O II] $\lambda\lambda$7323,7332 fluxes, respectively. In particular, $\simeq65\%$ of [O III] $\lambda$4363 comes from models with $\log(n_{\mathrm{e}}/\mathrm{cm^{-3}})>4$. The fact that nebular and auroral lines are emitted by different gas components can introduce significant biases in abundance estimates when assuming a uniform density traced by, for instance, the [S II] doublet alone.
We return to this issue in Sec. \ref{sec:marta_impact}.

The physical and chemical properties of the best-fit models are summarised in Table \ref{tab:Marta properties}. While not identical to the values presented in \citet{Curti2025}, they are generally in broad agreement.

The best-fit $A_{\mathrm{V}}=0.75^{+0.08}_{-0.02}$ mag is qualitatively in agreement (within $2\sigma$) with the value of $A_{\mathrm{V}}=0.496\pm 0.016$ mag obtained by \citet{Curti2025} via a least-squares fit to all available hydrogen lines.
In fact, H$\delta$/H$\beta$ and H$\gamma$/H$\beta$ suggest lower extinction, but H$\alpha$/H$\beta$ is not consistent with such inference (see \citealt{Curti2025} for a more detailed discussion).
We also repeated the fit excluding H$\alpha$ finding no change in the inferred $A_{\mathrm{V}}$, indicating that helium lines also provided support for a higher attenuation.
On the other hand, for instance, removing from the fit Paschen lines first and then also all helium lines yields $A_{\mathrm{V}}=0.42 \text{ and }0.29$ mag, respectively.
This result highlights the importance of including multiple diagnostics rather than relying solely on Balmer or hydrogen recombination lines for accurate dust-attenuation measurements.

Although HOMERUN is not yet optimised to recover the exact parameters of the stellar continuum and thus it is not designed to constrain stellar population properties \citep{Marconi2024}, we report here the physical quantities defining the best-fit stellar contiuum.
For M~4327, the stellar age is consistent with a young population of 2.5 Myr and the metallicity is $\log(Z_{\star})=-2.7$.

\subsection{Impact of density fluctuations on $T_{\mathrm{e}}$ abundances}
\label{sec:marta_impact}

The left panel of Figure \ref{Abund M4327} shows the difference between the best-fit HOMERUN abundances (obtained with our three different models) and those computed via the $T_{\mathrm{e}}$-method applied to the observed fluxes.
In the low-density case (yellow diamonds), HOMERUN estimates agree well with the $T_{\mathrm{e}}$-method for all elements.

The discrepancy between HOMERUN and $T_{\mathrm{e}}$ metallicities in the higher-density models reflects the role of dense ($n_\mathrm{e} \gtrsim 10^3$ cm$^{-3}$) regions in abundance determinations when diagnostics with low critical densities are involved. For example, [O II] $\lambda\lambda$3727,3729 has a critical density of $\sim 10^3$ cm$^{-3}$, while [O II] $\lambda\lambda$7323,7332 has a much higher one, $\sim 10^6$ cm$^{-3}$.
Therefore, in the presence of dense clumps ($\log(n_{\mathrm{e}}/\mathrm{cm}^{-3}) \gtrsim 3$), collisional de-excitation becomes effective for the [O II] $\lambda\lambda$3727,3729 transition but not for its auroral counterpart, leading to artificially high [O II] $\lambda\lambda$7323,7332 / [O II] $\lambda\lambda$3727,3729 ratios at fixed metallicity.
This behaviour biases the inferred $T$(O II) high when the electron density is derived from diagnostics that predominantly trace low-density gas, such as [S II] $\lambda\lambda$6717,6731 \citep[e.g.,][]{Me2023}.


The right panel of Figure \ref{Abund M4327} compares $T_{\mathrm{e}}$-based abundances obtained from observed emission lines and from fluxes predicted by the best-fit model. In this case, no clear trend with the maximum density is observed.
In addition, although the observations are not always perfectly reproduced by the best-fit models (Fig. \ref{M4327 chisq}), all the derived abundances agree within $<0.1$ dex.
This indicates that the discrepancy between model and $T_{\mathrm{e}}$ abundances primarily originates from the intrinsic limitations of the classical method and not the inability of the model to reproduce the auroral lines.

In conclusion, significant biases in the oxygen abundance can arise even if a minor fraction of the gas is at very high density, without resulting in a high average density, as seen in Table \ref{tab:Marta properties}, where all the models have comparable median density. This high-density gas does not contribute significantly to the typical density tracers adopted in the optical ([S II] $\lambda\lambda$6717,6731). To overcome this issue, it is important to use different diagnostics (e.g., UV lines) that probe a more ionised and dense zone of the ISM, especially in clumpy and compact environments such as those at high redshift (see Sec. \ref{UV diagnostic section}).
Indeed, from the best-fit model predictions for the C III] doublet we obtain $\log(n_{\mathrm{e}}(\mathrm{CIII]})/\mathrm{cm^{-3}})=4.6, \ 4.1$ in the two high-density cases, showing that UV diagnostics are sensitive to the presence of even a small amount of high-density material.




\begin{figure*}[ht]
    \centering
    \includegraphics[width=\linewidth]{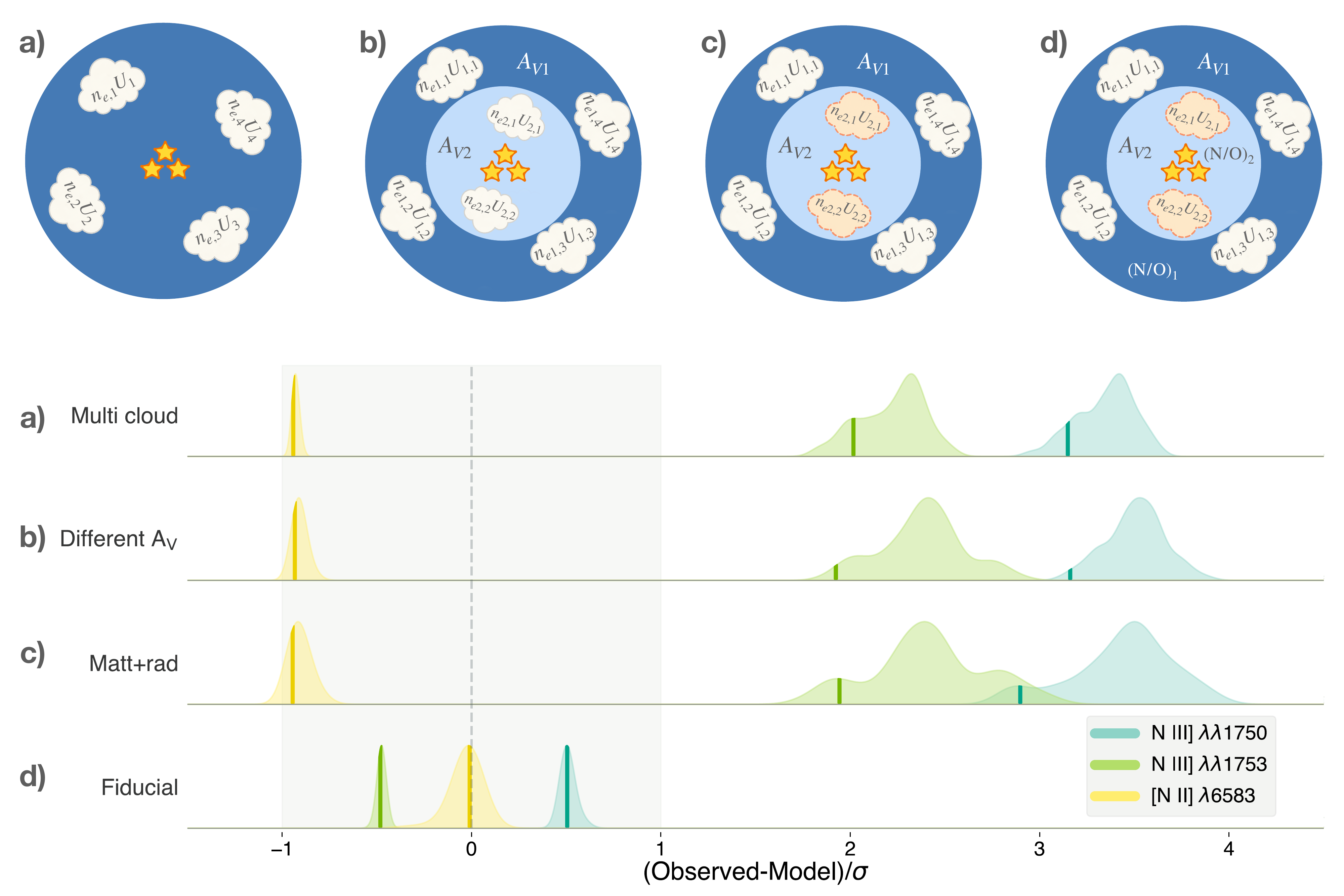}
    \caption{\textit{Top}: Cartoon illustrating the main assumptions of each JWST+MUSE model discussed in the text. This just serves as visual representation: HOMERUN does not provide any information about the actual relative spatial distribution of the models. \textit{Bottom}: Performance of different modelling approaches in reproducing nitrogen optical and UV fluxes in the Sunburst Arc. The probability distributions are obtained considering all models with $\mathcal{L} \leq \mathcal{L}_{\mathrm{min}} + 0.25$, smoothing the distributions to aid visualisation. The vertical coloured lines mark the best-fit value for each considered feature, and the grey shaded area the acceptable range as described in Sec. \ref{subsec:grids of cloudy models}. The multi-cloud model represents a typical HOMERUN model. The other three models consider two components with increasing levels of freedom. In the second row we allow the two components to have different $A_{\mathrm{V}}$. In the third row we explore the possibility of a combination of matter- and ionisation-bounded components. Finally, the bottom row shows the results of the fiducial model, where we consider a matter-bounded and radiation-bounded component with different attenuation and different N/O.}
    \label{fig:sunburst_nlines}
\end{figure*}
\section{Chemical stratification in the Sunburst Arc} \label{Sunburst models}
The Sunburst Arc is a highly magnified galaxy at redshift $z=2.37$, first identified by \citet{Dahle2016}.
In this work, we focus on the LyC-emitting region hosting a super star cluster with signatures of WR stars and possibly VMS \citep[e.g.,][]{RiveraThorsen2017, Mevs2023, RiveraThorsen2024}.
For simplicity, we refer to this system as the Sunburst Arc throughout the paper.
In our analysis, we use the rest-frame optical JWST fluxes from \citet{Welch2025}, obtained with the  NIRSpec Integral Field Unit (GO-2555, PI: Rivera-Thorsen) with two grating/filter settings (G140H/F100LP and G235H/
F170LP), and the rest-frame UV fluxes from MUSE as reported in \citet{Pascale2023} (programme 107.22SK, PI: Vanzella).

\subsection{Modelling the rest-frame optical lines (JWST model)}

We develop new photoionisation models for the Sunburst Arc using HOMERUN and the rest-frame optical JWST fluxes reported in \citet{Welch2025}, excluding [S III] $\lambda$6312 (the reasons for this choice are detailed in Appendix \ref{app:app:B1}). The inferred physical parameters for this `Sunburst-JWST' model are reported in Table \ref{tab:Sunburst properties}. In this case, even allowing the single-cloud models with densities up to $10^7 \ \mathrm{cm^{-3}}$, we infer a best-fit metallicity of $12+\log(\text{O/H}) = 7.97$ consistent with $12+\log(\text{O/H})=7.95^{+0.16}_{-0.13}$ we obtained from $T_{\mathrm{e}}$-method.
Even though uncertainties on HOMERUN metallicity are up to $\simeq0.3$ dex, this agreement demonstrates that model-based metallicities are not always significantly higher than those derived from the $T_{\mathrm{e}}$ method.
However, this is a special case, where even the [S II] doublet yields a high density ($n_\mathrm{e}\sim 10^3$ cm$^{-3}$).


\subsection{UV \& optical model (JWST+MUSE model)} \label{sec: Sunburst fiducial model}

We combined the JWST fluxes with the rest-frame UV MUSE fluxes from \citet{Pascale2023}. To minimise systematics in the absolute flux levels (due to potential differences in aperture and calibration), we normalised MUSE fluxes to H$\beta$ from X-shooter's NIR channel, since the X-shooter fluxes match the MUSE ones in the optical \citep{Pascale2023}. JWST fluxes were scaled using their own H$\beta$ line and an additional free scaling factor between the two datasets was included in the HOMERUN fit. The value for this scaling factor obtained for our fiducial best-fit is 0.87. 
We verified that excluding this factor only slightly worsens the agreement between observed and predicted fluxes, without impacting the derived physical properties, and thus the robustness of our results. To compare more directly with the analysis in \citet{Pascale2023}, we adopted the \citet{Calzetti2000} extinction law, with $A_{\mathrm{V}}$ left as a free parameter during the fit.

As done for the `Sunburst-JWST' model we first tested a simple multi-cloud HOMERUN fit.
However, this model fails to reproduce all the observed lines. In particular, it severely underestimates the N III] $\lambda\lambda$1750,1753 lines in the rest-frame UV by factors of $2-3$, while slightly overpredicting optical [N II] $\lambda$6583 ($<10\%$). Figure \ref{fig:sunburst_nlines}a shows the posteriors of the predicted nitrogen lines for this model, highlighting the large discrepancy of up to $0.3-0.4$ dex for N III] transitions. This failure shows that the optical and UV emission lines of nitrogen are mutually incompatible.

We therefore constructed increasingly complex HOMERUN models as a combination of two different components. Each individual component consists of a multi-cloud model with the same ionising continuum and oxygen abundance, but we relax some of the assumptions we previously made such as the uniformity in extinction or chemical abundances between the two components. 
As a first test, we consider a two-component model where we allow each component to have a different dust attenuation. This is meant to test whether differential attenuation can solve the UV-optical inconsistency in the nitrogen lines. 
Such an initial guess is reasonable given the difference in attenuation inferred from optical (Balmer) lines and UV lines in blue compact dwarfs, as well as the expectation that stars of different ages may be born in environments with different dust content \citep[e.g.,][]{Fanelli1988, Charlot2000}.
However, the attenuation is well-constrained by a variety of emission lines in the optical, and increasing the attenuation for the component mostly emitting in the UV does not resolve the tension, since these lines are already too faint in the model (Figure \ref{fig:sunburst_nlines}b). 
We then allowed one component to be matter-bounded.
Such matter-bounded regions would be plausible sources of LyC photon escape, a known feature of the Sunburst Arc \citep[e.g.,][]{RiveraThorsen2017}, and could enhance emission from high-ionisation species \citep[e.g.,][]{Brinchmann2008, Plat2019, McClymont2025a}.
In addition, a matter-bounded scenario is sometimes considered to interpret the anomalous Balmer decrements observed at high redshift \citep[e.g.,][]{Scarlata2024, Yanagisawa2024a,McClymont2025a}
This model, however, also fails in reconciling the various nitrogen lines (Figure \ref{fig:sunburst_nlines}c, third row).

Finally, we introduced chemical stratification by allowing N/O ratios (and $A_{\mathrm{V}}$) to differ between the two components, following the approach of \citet{Pascale2023}.
Interestingly, simply assuming that the Sunburst Arc is composed of two components with distinct N/O ratios and independent $A_{\mathrm{V}}$ values (without imposing a priori any specific high- or low-density zones or fixing the values of the N/O abundances) leads HOMERUN to fit both optical and UV lines within 10\% or within uncertainties (Figure \ref{fig:sunburst_nlines}d). While this scenario is consistent with that of \citet{Pascale2023},  we are able to simultaneously reproduce more lines with better accuracy and a general approach.
This model is also the most successful in terms of having the lowest loss. In particular, we find that component 1 has high $\log U$ ($\log U = -1.38$), high density ($\log(n_{\mathrm{e}}/ \mathrm{cm^{-3}}) = 4.5$), high N/O ($\log(\mathrm{N/O}) = -0.46$) compared to the value typical of its metallicity, negligible extinction ($A_\mathrm{V} = 0.00$), and a preference for matter- rather than ionisation-bounded clouds. Component 2, on the other hand, consists of more `typical' clouds, with lower ionisation parameter ($\log U = -1.96$), lower densities ($\log(n_{\mathrm{e}} / \mathrm{cm^{-3}}) = 2.0$) and an N/O ratio typical of low-metallicity systems ($\log(\mathrm{N/O}) = -1.36$).
We consider this our fiducial JWST+MUSE model for the Sunburst Arc and discuss further its implications in Sec. \ref{sec:stratification}.
The physical parameters of the best-fit solution for each model are reported in Table \ref{tab:Sunburst properties}.

\begin{table}
\caption{HOMERUN ISM properties for the Sunburst Arc.}
\centering
\renewcommand{\arraystretch}{1.3}
\begin{tabular}{lcc}
\toprule
\midrule
&  & Fiducial \\
& JWST-only & JWST+MUSE \\
\midrule
$<\log U_{\text{tot}}>$ & $-2.09^{+0.21}_{-0.01}$ & $-1.82^{+0.03}_{-0.28}$ \\
$<\log (n_\mathrm{e, tot} /\mathrm{cm}^{-3})>$ & $3.55^{+0.16}_{-0.33}$ & $2.55^{+0.71}_{-0.08}$ \\
$<\log U_1>$ & – & $-1.38_{-0.07}^{+0.00}$ \\
$<\log U_2>$ & – & $-1.96^{-0.08}_{-0.37}$ \\
$<\log (n_\mathrm{e, 1}/\mathrm{cm}^{-3})>$ & – & $4.49^{+0.07}_{-0.07}$ \\
$<\log (n_\mathrm{e, 2}/\mathrm{cm}^{-3})>$ & – & $1.95^{+0.89}_{-0.08}$ \\
$12+\log(\mathrm{O}/\mathrm{H})$ & $7.97^{+0.28}_{-0.06}$ & $8.19^{+0.11}_{-0.09}$ \\
$\log(\mathrm{N}/\mathrm{O})_{\text{tot}}$ & $-0.97^{+0.02}_{-0.11}$ & $-1.14^{+0.09}_{-0.06}$ \\
$\log(\mathrm{N}/\mathrm{O})_1$ & – & $-0.46^{+0.17}_{-0.07}$ \\
$\log(\mathrm{N}/\mathrm{O})_2$ & – & $-1.36^{+0.06}_{-0.33}$ \\
A$_V$& $0.34^{+0.07}_{-0.09}$ & $0.73^{+0.06}_{-0.03}$ \\
A$_{V,1}$& -- & $0.00^{+0.00}_{-0.00}$ \\
A$_{V,2}$& -- & $0.966^{+0.073}_{-0.013}$ \\
\midrule
\bottomrule
\end{tabular}
\tablefoot{HOMERUN gas-phase abundances and properties for the Sunburst Arc for both the model using only rest-frame optical line fluxes from JWST (`JWST-only') and the fiducial model using JWST+MUSE line fluxes which includes two components with different $A_{\mathrm{V}}$ and N/O, shown with the subscripts 1 and 2. In this latter model, component 1 is high-density and N-enriched, while component 2 is lower density and has a typical N/O ratio for its metallicity. For the JWST+MUSE model we also quote the flux-weighted mean $\logU$, $n_\mathrm{e}$ and N/O with the subscript tot. We only report here the chemical abundances discussed in the main text, and list the others in Appendix \ref{app:AppD}.}
\label{tab:Sunburst properties}
\end{table}
In the case of the Suburst Arc the best-fit stellar continuum model corresponds to an age of 5 Myr and stellar metallicity $\log(Z_{\star})=-2.7$. Both these values show good agreement with those ($t_\mathrm{age}\simeq4$ Myr and $\log(Z_{\star})\simeq-2.4$) found by \citet{RiveraThorsen2024} comparing the relative strengths of He II $\lambda$4686 and C IV $\lambda$5808 bumps with BPASS models, and also with estimates from other methods \citep{Chisholm2019, Pascale2023}.

\section{A nitrogen-enriched system at reionisation: RXCJ2248-ID}
\label{sec:Topping_models}

Until now, we have modelled systems at Cosmic Noon, while here we consider a lensed galaxy at redshift $z\sim6.11$. 
First identified by \cite{Boone2013}, RXCJ2248-ID shows high-ionisation lines like C IV $\lambda\lambda$1548,1551, He II $\lambda$1640, N IV] $\lambda\lambda$1483,1486, a low-metallicity and kinematically complex, dense and highly density stratified ISM ($n_\mathrm{e}\sim500-10^5 \ \mathrm{cm^{-3}}$; \citealt{Topping2024, CrespoGo2025, Berg2025}).
Although UV-diagnostic line ratios are similar with that of galaxies at $z\sim10$ \citep[e.g., GHZ2; ][]{Topping2024, Castellano2024}, the lower redshift of RXCJ2248-ID allows access to a wider range of spectral features. 
The recent detection of WR features by \citet{Berg2025} and comparison with constant pressure models \citep{Zhu2025b} favour a pure-SF scenario, but a contribution to ionisation from an AGN cannot be ruled out based on UV diagnostic diagrams \citep[e.g., ][]{Topping2024}. Thus, we also explore this latter possibility.

JWST observations discussed in our work were performed with the NIRSpec MSA using G140M/F100LP, G235M/F170LP, and G395M/F290LP gratings and filters under the programme GO-2478 (PI: Stark) \citep{Topping2024}.
The new JWST data presented by \citet{Berg2025} have higher sensitivity, but only cover the rest-frame optical range. In addition, the authors do not publish the fluxes for all detected emission lines. 
To maximise the number of features included in our models, we decided to adopt the fluxes from \citet{Topping2024}, which, in most cases, still agree with new measurements within our fiducial 10\% threshold.
Moreover, we also add [S II] $\lambda\lambda$6717,6731/H$\beta$ and [Ar IV] $\lambda$ 4711/H$\beta$ from \citet{Berg2025} to derive S/O and Ar/O.


\subsection{Photoionisation modelling with various ionising sources} \label{sec: Topping multi comp mods}

We tested a simple multi-cloud HOMERUN model and composite models combining (radiation-bounded) stellar and AGN continua, matter- and radiation-bounded stellar models, and finally a case with the same physical setup as the latter but allowing N/O to vary independently in the two components.
Indeed, the N/O abundance derived in the literature using the $T_{\mathrm{e}}$-method from optical differs from that derived by the UV lines generally by at least $\simeq0.2$ dex \citep{Topping2024, ArellanoCo2025, Martinez2025,Berg2025}.

In contrast to the Sunburst Arc, where a multi-cloud HOMERUN model struggled to reproduce UV and optical nitrogen lines simultaneously, in RXCJ2248-ID this simple model fails to reproduce high-ionisation species such as He II $\lambda$1640 and C IV $\lambda\lambda$1548,1551.
By introducing a harder ionising continuum such as that from an AGN or accounting for matter-bounded regions that reveal emission from deeper nebular zones, the agreement improves significantly.
However, He II $\lambda$1640 and C IV $\lambda\lambda$1548,1551 are complex transitions, both suffering from absorption, resonant scattering and possible contributions from stellar-wind features \citep[see also ][]{Zhu2025b}. The presence of WR stars in RXCJ2248-ID may also explain with stellar origin part of the observed flux of He II. However, at low metallicities ($12+\log(\mathrm{O/H})\lesssim8.1$) the nebular emission in C IV and He II usually dominates over the stellar one \citep{Senchyna2017}.

Based on a comparison with the UV diagnostic diagrams and models from \citet{Feltre2016} and \citet{Gutkin2016}, \citet{Topping2024} classified this galaxy as predominantly star forming. However, they also noted that a mixed contribution from star formation and AGN could not be ruled out. Moreover, the UV diagnostic diagrams in \citet{Hirschmann2019} place RXCJ2248-ID in the AGN+SF composite region.
New and deeper observations from the GLIMPSE survey allowed \citet{Berg2025} to identify WR star features in RXCJ2248-ID. The presence of WR stellar winds is also consistent with the complex kinematic and broadening of the observed emission lines.
We therefore take our pure SF model without chemical stratification ('Matter+Rad') as fiducial, even if a minor AGN contribution ($\simeq6$\% of H$\beta$) cannot be excluded based solely on the detected emission lines.

In fact, the three composite models described above are nearly degenerate, with $\mathcal{L}_{\min}$ differing by $\lesssim0.2$. 
They all systematically underpredict He I $\lambda$4471 multiplet but still within $2\sigma$, similarly to results by \citet{Yanagisawa2024}, who employed a modified MCMC approach to model H, He, and N lines.
However, the underlying absorption and radiative transfer effects can have a significant impact on helium lines.
Indeed, the fluxes reported by \citet{Berg2025} for this transition are more in agreement with our measurements.
Allowing different N/O ratios in the two star-forming components improves the fit, though not significantly.
Therefore, unlike the Sunburst Arc, chemical stratification is not required here, although it cannot be excluded. The model including stratification suggests a high-density, N-enriched zone coexisting with a low-density, N-normal one.

Table \ref{tab:Topping properties} summarises the best-fit results for the models tested. 
Overall, average ISM properties are consistent across configurations, even though they might be different for the single components.
Metallicity is significantly higher than the value estimated via the $T_{\mathrm{e}}$-method under standard assumptions, as we will discuss in the next section. 
The agreement between predicted and observed fluxes for each best-fit model is shown in Appendix \ref{app:AppC}.


Qualitatively in agreement with \citet{Topping2024}, our best-fit star-forming continuum corresponds to a stellar metallicity of $\sim10^{-2} \ \mathrm{Z}_{\odot}$ and young age of 1 Myr. For the AGN+SF model, the best-fit continuum instead yields age 12 Myr and, again, metallicity $10^{-2}$ Z$_{\odot}$, while the AGN model is defined by $\alpha_{\mathrm{ox}}=-1.2$ and an effective temperature of $T_\mathrm{max}=10^{5.5}$ K.




\begin{table}
\caption{HOMERUN ISM properties for RXCJ2248-ID.}
\centering
\renewcommand{\arraystretch}{1.3}
\begin{tabular}{lccc}
\toprule
\midrule
& \multicolumn{1}{c}{Fiducial} \\
& \multicolumn{1}{c}{Matter+Rad} 
& \multicolumn{1}{c}{AGN+SF} 
& \multicolumn{1}{c}{Free N/O} \\
\midrule
$<\log U_{\text{tot}}>$  
& $-1.25^{+0.08}_{-0.32}$ 
& $-1.01^{+0.13}_{-0.27}$ 
& $-1.44^{+0.41}_{-0.09}$ \\

$<\log (n_\mathrm{tot}/\mathrm{cm}^{-3})>$  
& $3.32^{+1.12}_{-0.09}$ 
& $2.9^{+0.2}_{-0.2}$ 
& $3.86^{+0.16}_{-0.75}$ \\

$<\log U_1>$  
& $-1.50^{+0.50}_{-0.00}$ 
& $-0.59^{+0.91}_{-0.10}$ 
& $-1.74^{+0.54}_{-0.12}$ \\

$<\log U_2>$  
& $-1.22^{+0.08}_{-0.37}$ 
& $-1.04^{+0.00}_{-0.54}$ 
& $-1.24^{+0.24}_{-0.04}$ \\

$<\log (n_\mathrm{e, 1}/\mathrm{cm}^{-3})>$  
& $7.00^{+0.00}_{-1.00}$ 
& $2.83^{+0.93}_{-0.12}$ 
& $5.5^{+0.9}_{-0.5}$ \\

$<\log (n_\mathrm{e, 2}/\mathrm{cm}^{-3})>$  
& $2.81^{+1.17}_{-0.07}$ 
& $4.63^{+0.00}_{-2.73}$ 
& $2.87^{+0.08}_{-0.78}$ \\

$12+\log(\mathrm{O}/\mathrm{H})$  
& $7.81^{+0.15}_{-0.13}$ 
& $7.81^{+0.11}_{-0.11}$ 
& $7.87^{+0.08}_{-0.16}$ \\

$\log(\mathrm{N}/\mathrm{O})_{\text{tot}}$ 
& $-0.55^{+0.04}_{-0.02}$ 
& $-0.54^{+0.04}_{-0.03}$ 
& $-0.76^{+0.25}_{-0.08}$ \\

$\log(\mathrm{N}/\mathrm{O})_1$ 
& - 
& - 
& $-1.35^{+1.01}_{-0.04}$ \\

$\log(\mathrm{N}/\mathrm{O})_2$ 
& - 
& - 
& $-0.35^{+0.06}_{-0.33}$ \\

$\mathcal{L}_\mathrm{min}$ 
& 0.679 
& 0.499 
& 0.642 \\
\midrule
\bottomrule
\end{tabular}

\tablefoot{In these composite models, subscripts 1 and 2 refer to the matter-bounded and ionisation-bounded component, or to AGN and SF in the corresponding scenario. We do not report A$_V$, since it is identically 0 for all our models. Additional elemental abundances are reported in Appendix \ref{app:AppD}.}
\label{tab:Topping properties}
\end{table}

\section{Discussion}
\label{sec:discussion}
\subsection{The impact of density and temperature fluctuations}
We have quantitively demonstrated in this work that the two-zone approximation underlying the $T_{\mathrm{e}}$-method is a poor assumption when compared to the complexity of real galaxies and HII regions, where emission may arise from zones with significantly different ionisation parameters and densities.
In the following, we discuss the capabilities and potential biases of standard optical and UV tracers in the presence of such structures.
\begin{figure*}[ht]
    \centering
    \includegraphics[width=\linewidth]{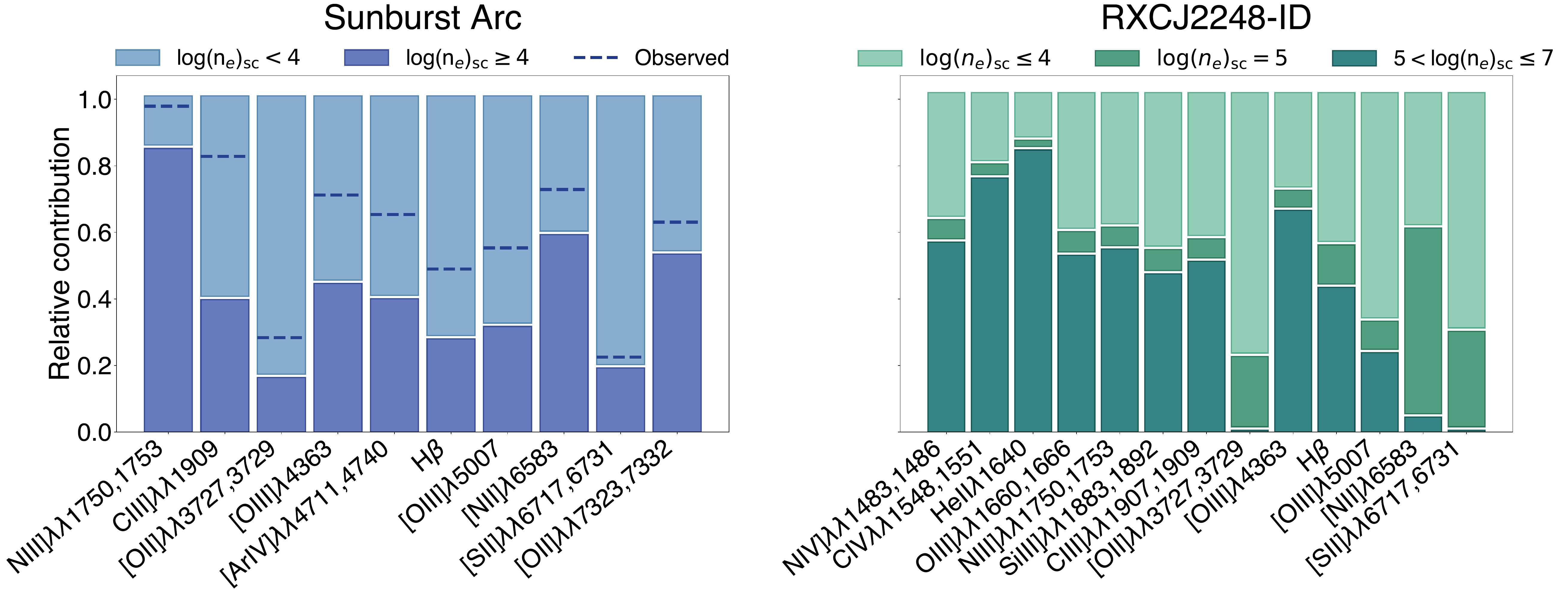}
    \caption{\textit{Left}: analogue to Figure \ref{MARTA 4327 rel contr} but for the Sunburst Arc and separating the contribution from matter- and radiation-bounded components. Since in this case the two component in our best-fit have different attenuation, we show as dashed lines the "observed" relative contribution (i.e., reddening the fluxes in CLOUDY accordingly to the best-fit A$_V$ reported in Table \ref{tab:Sunburst properties}). \textit{Right}: similar for RXCJ2248-ID and assuming the Matter+Rad case. We also distinguish the contributions from single-cloud models with $\log(n_{\mathrm{e}}/\mathrm{cm}^{-3})=5$ and $\log(n_{\mathrm{e}}/\mathrm{cm}^{-3})=6,7$.}
    \label{Relative contr}
\end{figure*}

\subsubsection{The limitations of optical diagnostics} 
\label{sec:limitations of opt diagn}


The complexity introduced by density and temperature inhomogeneities poses concrete challenges for traditional optical diagnostics.
Density inhomogeneities primarily affect the low-ionisation zone, where diagnostics such as the [O II] and [S II] doublets (which have low critical densities) are most sensitive.
The [N II] $\lambda5755/\mathrm{[N \ II]} \lambda6583$ auroral-to-nebular ratio, instead, is largely insensitive to density variations up to $n_\mathrm{e} \sim 10^5$ cm$^{-3}$ and (unlike [O II]) is not significantly affected by dielectronic recombination in dense ($n_\mathrm{e} \gtrsim 10^3$ cm$^{-3}$), moderately metal-rich environments \citep[e.g.,][]{Liu2001,Me2023, Welch2025}.
Consequently, whenever available, the nitrogen temperature diagnostic should be preferred over those based on [S II] or [O II] \citep[e.g.,][]{Berg2020, Zurita2021, Me2023a, Me2023}.
However, this remains observationally extremely rare in single galaxies at high redshift, even with JWST, due to the faintness of the [N II] auroral line, especially at the relatively low metallicities characteristic of such systems \citep{Morishita2025b, Rogers2025}.

Regarding the high-ionisation zone, [O III] $\lambda5007$ and $\lambda4363$ have critical densities of $\sim$ a few $10^5$ cm$^{-3}$ and $\sim10^7$ cm$^{-3}$, respectively. Therefore, their ratio remains largely unaffected by density unless extreme conditions with very dense ($\gtrsim10^5$ cm$^{-3}$) and highly ionised clumps are present, as encountered, for example, in RXCJ2248-ID (see Sec. \ref{UV diagnostic section}).

Temperature inhomogeneities further complicate abundance
estimates. Since auroral lines preferentially trace hotter gas, temperature measurements can be biased high in the presence of temperature fluctuations, leading to underestimated metallicities \citep[e.g.,][]{Peimbert1967, Stasin2005, Me2023a, Peng2025a}.
The classical formalism introduced by \citet{Peimbert1967} quantifies temperature fluctuations through the root-mean-square temperature-fluctuation parameter $t^2$ \citep{Peimbert1967}.
We compute $t^2$ with our multi-cloud approach. For each single-cloud model, we use \texttt{PyNeb} to derive $T$(O III) from [O III] $\lambda4363$ / [O III] $\lambda5007$ fixing the density to the model one. We then compute the average and mean-squared temperatures weighted by the best-fit model weights. With this approach we are effectively neglecting the temperature gradients inside each single-cloud model, which we assume to be sub-dominant. Doing so yields $t^2$ values of 10\%, 19\%, and 11\% for the fiducial models of M~4327, the Sunburst Arc, and RXCJ2248-ID, respectively.
We comment further on this topic in Appendix \ref{app:J}.

\subsubsection{The power of UV and multiple diagnostics} \label{UV diagnostic section}

Both the Sunburst Arc and RXCJ2248-ID have rest-frame UV coverage and a detection of the density-sensitive C III] $\lambda\lambda$1907,1909 doublet, which has a critical density of $\sim 10^6$\,cm$^{-3}$. 
In the case of the Sunburst Arc, the metallicities derived from the HOMERUN best-fit models presented in Sec. \ref{Sunburst models} are $\simeq0.1-0.3$ dex lower than the values obtained via the $T_{\mathrm{e}}$-method (but all consistent within the uncertainties). While the [S II]-derived density is quite high ($\sim10^3$ cm$^{-3}$), from C III] $\lambda\lambda$1907,1909 we obtain $n_\mathrm{e}(\mathrm{C \ III]})=102000\pm16000 \ \mathrm{cm}^{-3}$. 
Indeed, using the $T_{\mathrm{e}}$-method but assuming a constant density $n_\mathrm{e}(\mathrm{C\ III]})=10^5$ cm$^{-3}$ yields $12+\log(\mathrm{O/H})=8.35$. The latter value is in better agreement with our fiducial JWST+MUSE fit predictions of 8.19, and also with the estimate by \citet{Martinez2025}, who incorporated the $T_{\mathrm{e}}$-method in a more realistic three-zone approach.


As shown in the left panel of Figure \ref{Relative contr}, most of the contribution to UV lines comes from higher-density single-cloud models, while lower-density gas dominates the optical emission (except for [N II] $\lambda$6583 and the auroral lines [O III] $\lambda$4363 and [O II] $\lambda\lambda$7323,7332).
In the fiducial model for the Sunburst Arc, all single-cloud models with $n_\mathrm{e} < 10^4$\,cm$^{-3}$ belong to the `N-normal' component, which has only a $\simeq$3\% contribution from higher-density regions. Thus, the division into high-/low-density is closely mirrored by the split between high/normal N/O.
Figure \ref{Relative contr} also indicates that when the emission comes from regions with different attenuation properties, such as in our fiducial model, the lines preferentially emitted by the denser but lower-extinction component (e.g., the UV lines) are even more prominent in the observed spectrum.

For RXCJ2248-ID, observation recently published in \cite{Berg2025} give access to multiple density tracers, spanning a large ionisation range (10 - 77 eV), including the [Ar IV] $\lambda\lambda$4711,40, [S II] $\lambda\lambda$6717,6731 doublets. UV lines imply the highest densities ($n_\mathrm{e} (\mathrm{N \ IV}]) =2.65 \times 10^5$\,cm$^{-3}$) while the optical lines probe densities about two order of magnitude lower. 
Fig. \ref{Relative contr} shows that for our fiducial 'Matter+Rad' model of RXCJ2248-ID most lines receive a substantial contribution from clouds with $n_\mathrm{e} \geq 10^5$\,cm$^{-3}$. In this regime, even the high-critical density N IV] ratio may not capture the full contribution of highest-density regions, since our best-fit model predicts that $\simeq60\%$ of its flux comes from single-cloud models with $\log(n_{\mathrm{e}}/\mathrm{cm^{-3}})\geq6$.

In such extreme environments, [O III] $\lambda$5007 undergoes collisional de-excitation ($n_\mathrm{crit} \sim 10^5$\,cm$^{-3}$), while the corresponding auroral line does not. As a result, the [O III] $\lambda$4363/[O III] $\lambda$5007 ratio -and hence the inferred temperature- can be artificially biased high unless the density is properly constrained. In fact, assuming the density implied by [S II] gives a $T\rm_e$-metallicity of $7.62\pm0.02$, while assuming the density traced by N IV] leads to a metallicity of $7.81\pm0.04$, much closer to the oxygen abundance of the fiducial HOMERUN model.

This finding is in line with the conclusion of \cite{Martinez2025}, who argue that it is necessary to use UV density diagnostics to determine $T\rm(O \ III)$ in high-density environments, though the true density may still be underestimated. We note that differences in the physical origin of the nebular and auroral [O III] lines have been directly observed in blue compact dwarfs at low redshift, such as, e.g., Mrk~996 \citep{Thuan1996, James2009, Telles2014}.

Finally, although we focused here primarily on the effects of high-density clumps on abundance determinations, combining JWST and ALMA data, \citet{Harikane2025a} used a two-zone model to argue that ignoring a low-density gas component (below the limit of sensitivity of the [S II] doublet) can lead to a systematic underestimation of metallicity by $\simeq 0.2 - 0.4$ dex in high-redshift galaxies. 
From a different perspective, considering the MEGATRON cosmological hydrodynamical simulation \citep{Katz2025}, \citet{Choustikov2025} showed that no-two main UV, optical, and FIR emission lines originate from gas with the same density and temperature. This result holds even though the simulation primarly samples relatively low densities
compared to those expected in highly compact star-forming regions.
Therefore, in order to constrain ISM properties from emission lines in a highly density-stratified medium, it is critical to access multiple diagnostics that probe both high- and low-density phases, and not only UV-optical features.


\subsection{Nitrogen abundances and chemical stratification}
\label{sec:stratification}

Both the Sunburst Arc and RXCJ2248-ID are classified as N-enhanced galaxies, i.e. having a higher nitrogen abundance than expected for their metallicity considering the $z=0$ trend. 
High nitrogen abundances in these systems could be associated with regions near massive stars, where stellar winds locally enrich the ISM, potentially creating chemical inhomogeneities. As a results, the extreme N/O derived from UV lines in N-emitters would be representative of these localised, compact, enriched clumps rather than the whole galaxy, a scenario invoked for the Sunburst Arc \citep{Pascale2023}. However, due to the faintness of the spectral features involved and the required spectral coverage to detect both N III] $\lambda$1750\footnote{Here we refer to all the components of N III quintuplet, at wavelengths 1746.82, 1748.65, 1749.67, 1752.16, 1753.99 $\AA$.}, N IV] $\lambda\lambda$1483,1486, and [N II] $\lambda$6583, only a few other cases in the literature allow the investigation of inhomogeneities in N/O at high redshift \citep{Ji2024, Berg2025}. 

In the local Universe, on the other hand, spatially resolved studies of  blue compact dwarfs allow probing chemical inhomogeneities in high-redshift analogues \citep{Kumari2018,Kumari2019, James2020}. For example, Mrk~996 and NGC~5253 exhibit regions with N/O enhancement ($\simeq0.5-1$ dex) at almost constant O/H \citep[e.g.,][]{James2009, Prujit2025}. Such N/O-rich regions are found around - but not necessarily coincident with - massive stellar clusters, which enrich in nitrogen (and sometimes helium) the surrounding medium via stellar winds \citep{Kumari2018}. \cite{Prujit2025} showed that such variations may also correlate with dust distribution and density structures on pc scales.

\begin{figure*}[ht]
    \centering
    \includegraphics[width=0.45\textwidth]{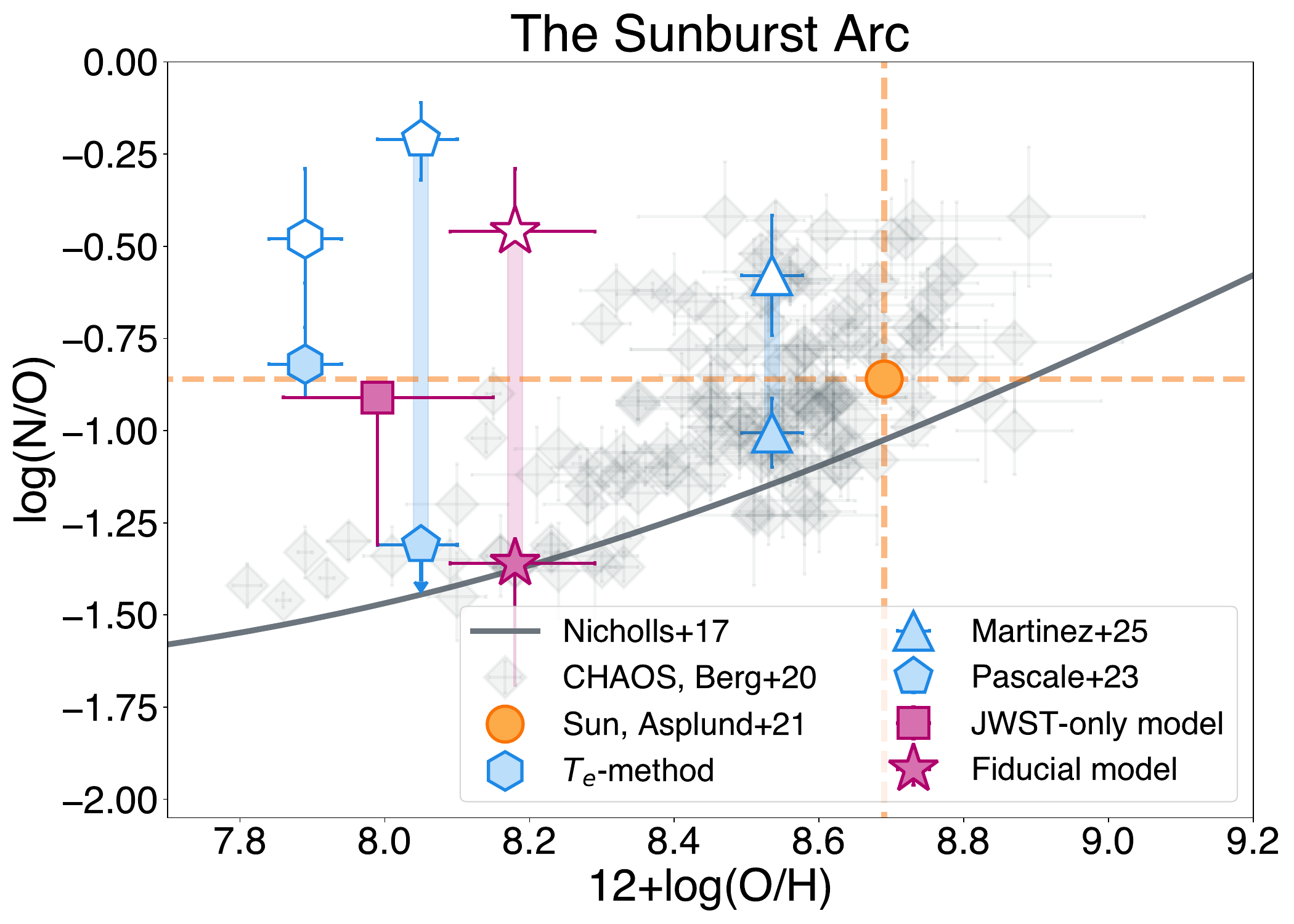}
    \hspace{0.05\textwidth}
    \includegraphics[width=0.45\textwidth]{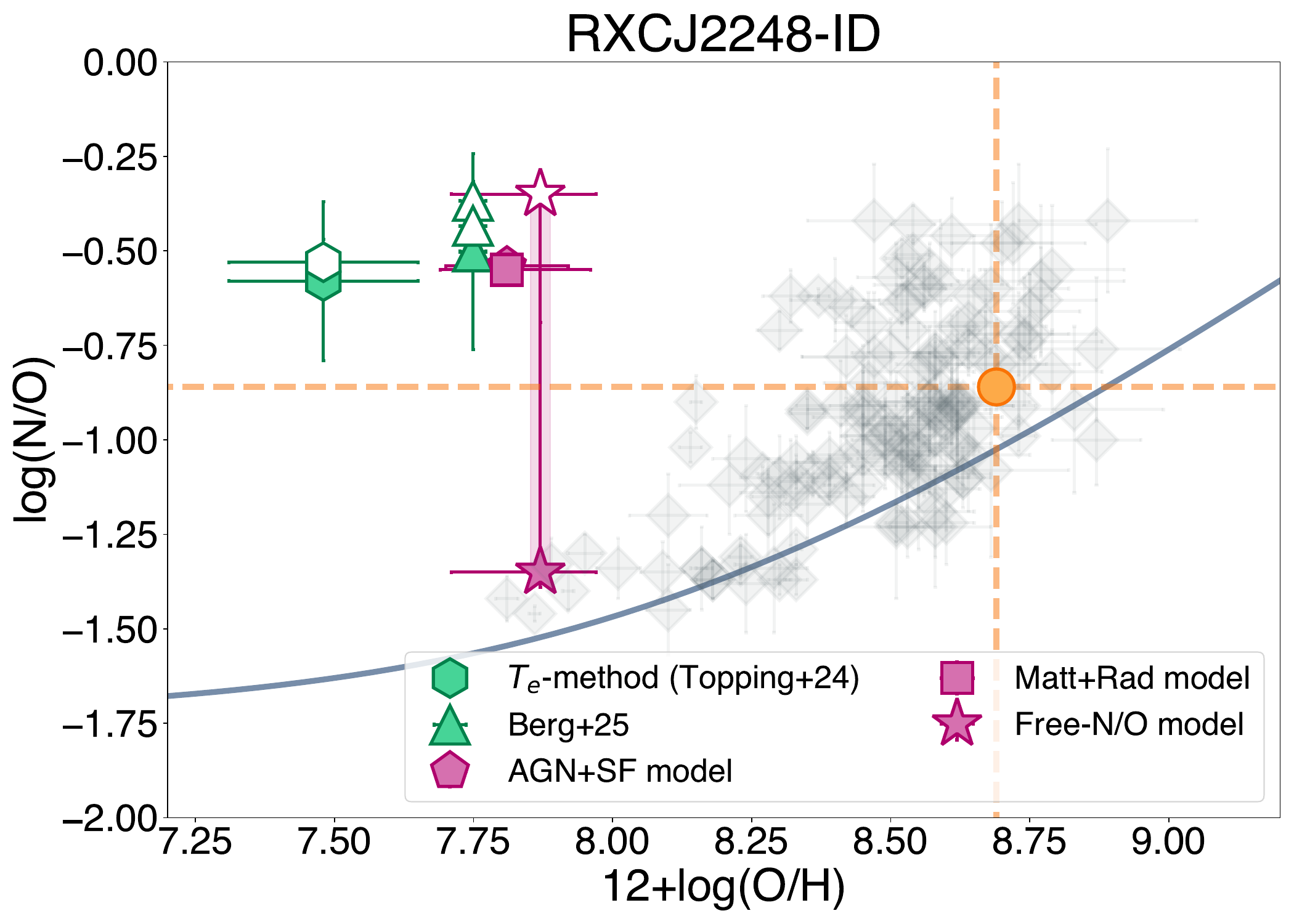}
    \caption{\textit{Left}: log(N/O) vs 12+log(O/H) for different estimates of abundances in the Sunburst Arc. The hexagon corresponds to the abundances obtained with $T_{\mathrm{e}}$-method and fluxes, the square to HOMERUN-abundance from the best-fit model to JWST-only fluxes. On the other hand, the pentagon shows abundances from the photoionisation model by \citet{Pascale2023} for both the low- and high-pressure zone (void marker for the latter), connected by the shaded area of the same colour. Similarly, the measurements for our fiducial JWST+MUSE model assuming chemical stratification are indicated by the purple stars, with the (void) filled marker corresponding to the (high-) low-density (matter-) radiation-bounded component. We also report the values obtained with a three-zone approach by \citet{Martinez2025} for UV (void triangle) and optical (filled triangle) N/O.  Also local extragalactic HII regions from \citet{Berg2020} (grey diamonds) and the stellar-based relation by \citet{Nicholls2017} (grey solid line) are reported as reference. The orange bullet and dashed lines represent the solar value by \citet{Asplund2021}. \textit{Right}: similar but for RXCJ2248-ID. Again, our best-fit models are the purple markers. In this case, the hexagons refer to $T_\mathrm{e}$-abundances recalculated with fluxes in \citet{Topping2024} (see also Appendix \ref{app:AppA}), triangles show the N/O obtained by \citet{Berg2025} with a three-zone approach and prescriptions from \citet{Martinez2025}.}
    \label{NO}
\end{figure*}

Figure \ref{NO}, left panel shows the position of the Sunburst Arc in the O/H versus N/O plane with abundances obtained using different methods or modelling assumptions.
Our fiducial model is consistent with the two-zone model by \citet{Pascale2023}, although our N/O in the high-density zone is $\simeq0.2$ dex lower. Notably, the fiducial model predicts the N/O of the low-density component to be consistent with the O/H-N/O relation from low-redshift data \citep[e.g.,][]{Nicholls2017}.
The model using only JWST fluxes yields a nitrogen abundance of $\log(\mathrm{N}/\mathrm{O}) = -0.97$, about 0.12 dex lower than the value we obtained via the $T_{\mathrm{e}}$-method, though still consistent within the uncertainties.
The optical lines themselves are not representative of either component: [O II] $\lambda\lambda$3727,2729 originates mainly in the low-density, N-normal zone, while [N II] $\lambda$6583 receives a significant contribution from the high-density, N-rich region (Figure \ref{Relative contr}, left panel). This behaviour leads the JWST-only model to N/O values that are intermediate between the ones of the two components in the chemically stratified model.
We also show in the figure the abundances inferred by  \citet{Martinez2025}, who analysed the effects of high-density inhomogeneities on metallicity and N/O measurements in a sample of local and high-redshift galaxies with multiple UV and optical density diagnostics, and found a  metallicity 0.3 dex higher than our fiducial model.

The right panel of Fig. \ref{NO} shows the abundance estimates for RXCJ2248-ID in the O/H versus N/O plane. First, we note that, assuming a typical $n_\mathrm{e}([\mathrm{S \ II]})=500 \ \mathrm{cm^{-3}}$ density for galaxies at $z=6$ \citep{Topping2025}, we would obtain $\log(\mathrm{N}/\mathrm{O})_{\mathrm{optical}} = 0.21$, which would be unphysically high (beyond the range of the plot in Fig. \ref{NO}). \cite{Martinez2025} reached a similar conclusion and argue therefore for taking the UV density tracers into account. In fact, using instead the density of the UV high-ionisation region ($n_\mathrm{e} = 10^5$\,cm$^{-3}$) yields $\rm log(N/O) = -0.56$, similar to the value obtained by \cite{Berg2025} and a striking 0.77 dex lower than the value obtained ignoring the high-density component.

Our fiducial `Matter+Rad' model for RXCJ2248-ID gives  $\log(\mathrm{N}/\mathrm{O}) = -0.55^{+0.04}_{-0.02}$, marginally consistent with the N/O values from \cite{Berg2025}.
In contrast to the Sunburst Arc, no chemical stratification is required to reproduce the observed line ratios in this galaxy. In fact, the difference between optical and UV-derived N/O can be explained by a structure in density and ionisation, in agreement with the conclusions of \cite{Berg2025}.
Nevertheless, the chemically stratified model cannot be ruled out, as discussed in Sec. \ref{sec:Topping_models}.
In this case, similarly to the Sunburst Arc, the best-fit model includes a high-density, nitrogen-enriched zone with $\log(\mathrm{N}/\mathrm{O}) = -0.35$, and a lower-density, nitrogen-normal zone with $\log(\mathrm{N}/\mathrm{O}) = -1.35$, remarkably consistent with the primary nitrogen plateau observed in local galaxies.
As before, the presence of high-density clumps biases the N/O inferred from optical lines toward this higher value: while [O II] $\lambda\lambda$3727,3729 mostly originates in the lower-density region, [N II] $\lambda$6583 -with its higher critical density- probes also denser gas (with roughly half of the flux coming from regions with density 10$^5$ cm$^{-3}$).
In these conditions, [O II] and [N II] trace different volumes, and thus their ratio may no longer be representative of the true N/O, especially in the case of chemical stratification.

\subsection{High-ionisation lines: harder spectrum or matter-bounded clouds?}

Disentangling a mixture of AGN and star-formation activity from the ionisation of massive metal-poor stars becomes increasingly challenging at higher redshift, where only brightest emission lines are detected. 
For example, RXCJ2248-ID shares similar spectral properties with $z\sim10$ sources such as GN-z11 and, in particular, GHZ2 \citep[][]{Topping2024, Castellano2024}.
With a HOMERUN NIRSpec+MIRI+ALMA combined analysis of GHZ2, \citet{Castellano2025} showed that the observed emission could be explained by both AGN+SF and matter+radiation bounded composite models. However, following the discussion in \citet{Castellano2025} the AGN+SF scenario is preferred, in agreement with the independent analysis of \citet{ChavezOrtiz2025} using BEAGLE AGN.

An AGN is not required to explain the observed emission-line spectrum in RXCJ2248-ID, if a matter-bounded contribution is included. Such a component can “mimic” the presence of a harder ionising spectrum, since matter-bounded nebulae expose the inner, high-energy regions of the ionised gas.
The existence of a matter-bounded component is also consistent with the upper limit on Mg II $\lambda\lambda$2797,2803, which suggests a low column density or covering fraction, and thus a non-negligible LyC escape \citep{Topping2024}. This interpretation could be further supported by the high C IV/C III] ratio \citep{Schaerer2022} and the extremely blue UV slope observed: as \citet{Cullen2025} note, the slope would be heavily reddened in fully ionisation-bounded clouds with $n_\mathrm{e}\gtrsim 10^4$ cm$^{-3}$ due to strong nebular-continuum emission. This effect could be mitigated in the presence of LyC escape \citep{Topping2024}.
Additionally, RXCJ2248-ID shows Balmer line ratios that deviate from Case B predictions, aligning more with the expectations of a matter-bounded nebula \citep{Yanagisawa2024a}.

Our matter-bounded single-cloud models are extreme and constructed specifically to reproduce a high He II emission (Sec. \ref{subsec:grids of cloudy models}). He II lines are particularly difficult to model with standard stellar populations and photoionisation frameworks.
Consequently, strong He II lines are often attributed to alternative ionisation sources such as extremely metal-poor massive stars, X-ray binaries, fast shocks, or low-luminosity AGN \citep[e.g.,][]{ Senchyna2017, Chevallard2018, Plat2019, Senchyna2019, Lecroq2024}.
Despite the significant uncertainties associated with our assumptions, we computed escape fractions $f_\mathrm{esc}$ from our fiducial HOMERUN model fits, obtaining  $f_\mathrm{esc} = 12\%$ for RXCJ2248-ID.
Similarly, for the Sunburst Arc fiducial model we find $f_{\mathrm{esc}} = 43\%$, in qualitative agreement with previous estimates of $f_{\mathrm{esc}} \simeq 30\%$ \citep{RiveraThorsen2019}. 

In Appendix \ref{app:B2}, we briefly discuss how including VMS in our stellar models could impact our results.


\section{Conclusions}
\label{sec:conclusion}
We used our novel photoionisation modelling framework (HOMERUN, \citealt{Marconi2024}) to explore the impact of ISM complexity on nebular diagnostics at high redshift. To this end, we carried out an in-depth analysis of the physical and chemical properties of the ionised ISM in three benchmark objects: two systems at Cosmic Noon (M~4327 and the Sunburst Arc) and one reionisation-era galaxy (RXCJ2248-ID).
We demonstrated that HOMERUN allows us to self-consistently model and probe a wide range of physical conditions found in high-redshift systems within a unified framework, without relying on additional assumptions or empirical calibrations.  
We summarise the main conclusions below.

\begin{itemize}
  \item HOMERUN is able to reproduce all observed emission lines with great accuracy in the objects considered in this work, typically within 10\% or the observational uncertainties.  
  \item Although not dominant in their Balmer line emission, high-density regions can still contribute most of the auroral-line emission in lines such as [O III] $\lambda$4363 and [O II] $\lambda\lambda$7323,7332. As a result, standard $T_{\mathrm{e}}$-based metallicities can be biased low, even in high-S/N spectra, especially when using diagnostics involving low-critical-density transitions like [O II] $\lambda\lambda$3727,3729.
  In M~4327, our best-fit model suggests the presence of unresolved  dense ($\log(n_{\mathrm{e}}/\mathrm{cm}^{-3})\geq4$), highly ionised clumps based on the full emission-line spectrum, despite the absence of far-UV density tracers. Due to such structures, the $T_{\mathrm{e}}$-based metallicity and N/O for this galaxy may be underestimated by $\simeq0.10 - 0.25$ dex.  
  In the Sunburst Arc, modelling only the optical lines gave a model metallicity in good agreement with the $T_{\mathrm{e}}$ estimate, but including UV lines again led to a $\simeq0.3$ dex offset.  
  However, UV lines may not always be enough for the more extreme systems at higher redshift. 
  In a heavily density-stratified medium, the emission from low-density but highly ionised clumps could also affect the density inferred from N IV] $\lambda\lambda1483,1486$. As a consequence, in RXCJ2248-ID the best-fit model metallicity exceeds the $T_{\mathrm{e}}$ estimate by $\simeq0.05-0.15$ dex (depending on the considered scenario), even when using a multi-zone approach.
  \item Our UV-to-optical modelling of the Sunburst Arc requires chemical inhomogeneities to simultaneously reproduce [N II] $\lambda$6583 and N III] $\lambda\lambda$1750,1753. The best-fit model comprises a dense, highly ionised, matter-bounded region enriched in nitrogen, coexisting with a more diffuse, chemically “normal” component.  
  However, a discrepancy in N/O derived from optical and UV diagnostics does not necessarily imply chemical stratification. Chemical inhomogeneities are not needed to explain nitrogen optical and UV emission lines in RXCJ2248-ID, although they cannot be ruled out.  
  In both the Sunburst Arc and RXCJ2248-ID, the N/O derived from UV lines using the $T_{\mathrm{e}}$-method matches that of the N-rich component, while the optical-derived N/O is not representative of either the N-rich or the N-normal phase.  
  When [N II] $\lambda$6583 and [O II] $\lambda\lambda$3727,3729 are emitted from distinct zones, their ratio may be a poor tracer of the “true” N/O in a chemically stratified medium.
  \item In a pure star-formation scenario for RXCJ2248-ID, HOMERUN can only reproduce the full suite of observed emission lines by including matter-bounded clouds. Otherwise, high-ionisation features such as C IV $\lambda\lambda$1548,1551 and He II $\lambda$1640 are severely underpredicted.  
  Although the detection of WR signatures in recent observations supports the presence of a young, massive stellar population as the driver for the observed emission, a minor narrow-line-AGN contribution cannot be excluded based solely on line fluxes. This conclusion reflects the broader challenge of distinguishing AGN activity in metal-poor galaxies powered by massive-star formation in dense environments, especially when spectra are not deep enough to resolve faint WR features.
\end{itemize}

Our results underline the critical need for physically motivated, multi-zone, and chemically flexible tools to interpret the increasingly rich spectroscopic datasets provided by JWST.  
Simple assumptions of uniform density, ionisation structure, or chemical composition fail to capture the extreme conditions and complexity of high-redshift galaxies. HOMERUN provides a powerful framework to overcome these limitations, offering a unified approach toward a more complete understanding of the physical and chemical properties of the ISM across cosmic time.

\begin{acknowledgements}
FB,  FM, GC, EB, AM acknowledge support from the INAF Fundamental Astrophysics programme 2022, 2023 and 2024, in particular Mini Grant 2023 ``Quantitative Spectroscopy of Ionized Nebulae and Galaxies (QSING)'', Data Analysis Gran 2024 ''Accurate measurements of metallicity in galaxies with a new approach to photoionization modelling''. FM, FB, AM and IL acknowledge support from project PRIN-MUR 202223XDPZM "Prometeus", financed by the European Union -  Next Generation EU, Mission 4 Component 1 CUP B53D23004750006.
\end{acknowledgements}

\bibliography{refs}

\begin{appendix}

\section{Modelling assumptions and chemical abundance determinations}
\label{app:AppA}

Although oxygen is the primary coolant in HII regions, abundant elements like nitrogen, sulfur, and carbon can also influence ISM cooling and, consequently, the predicted fluxes of single-cloud models. The predicted flux in these models may no longer scale linearly with the abundance of the corresponding element for significant variations in its abundance. 
However, we verified that the final results does not change when adopting instead model grids with N and S abundances rescaled by $\pm0.6$ and $\pm0.3$ dex, respectively, and limiting the scaling factors to vary within small amounts (as in \citealt{Marconi2024}).

Similarly, these scaling factors can also effectively absorb deviations in depletion factors from those assumed in the models. 
For instance, we verified that the iron abundance inferred by combining depleted and undepleted models is consistent with that obtained using a single model and allowing the iron line flux to vary by up to $\pm1$ dex.
In Table \ref{tab:models} we list for reference all the parameters of the single-cloud models described in Sec. \ref{subsec:grids of cloudy models}.
\begin{table}
\centering
\caption{Model parameters adopted for the photoionisation calculations.}
\label{tab:models}
\begin{tabular}{l c}
\toprule
\multicolumn{2}{c}{HII-region models} \\
\midrule
log($t_{\mathrm{age}}$/yr) & 6.0, 6.4, 6.6, 7.1, 8 \\
log($Z_{\star}$) & $-1.7$, $-2.7$, $-4$ \\
log($Z_{\mathrm{gas}}$/Z$_{\odot}$) & $[-2, \ 0.4]^{a}$ , $[-3.3, -0.3]^{b}$\\
$\logU$ & $[-4, \ -1]$ \\
log($n_{\mathrm{e}}$/cm$^{-3}$) & $[0,\ 7]$ \\
\midrule[\heavyrulewidth] 
\multicolumn{2}{c}{AGN models} \\
\midrule
log($T_{\mathrm{max}}$/K) & $[4.0, \ 7.0]$ \\
$\alpha_{\mathrm{ox}}$ & $-1.2, \ -1.5 \ ,-1.8$ \\
$\alpha_{\mathrm{x}}$ & $-1$ \\
$\alpha_{\mathrm{uv}}$ & $-0.5$ \\
log($Z_{\mathrm{gas}}$/Z$_{\odot}$) & $[-2, \ 0.4]$ \\
log($n_{\mathrm{e}}$/cm$^{-3}$) & $[0, \ 7]$ \\
$\logU$ & $[-4, \ 0.5]$ \\
\bottomrule
\end{tabular}
\tablefoot{
\tablefoottext{a}{Gas metallicity for models with log($Z_{\star})=-1.7$, $-2.7$, with solar metallicity Z$_{\odot}=0.014$ corresponding to $-1.9$.} 
\tablefoottext{b}{Gas metallicity for models with log($Z_{\star})=-4$.}
}
\label{single-cloud propr}
\end{table}


For comparison, we computed abundances for each object with the $T_{\mathrm{e}}$-method following standard practice, and summarise the detailed below.
First, we corrected the observed fluxes for reddening 
as derived from the HOMERUN best-fit model, adopting the attenuation curve detailed in the text: \citet{Cardelli1989} for M~4327, and \citet{Calzetti2000} for the Sunburst Arc and RXJCJ2248-ID. 
To simplify the analysis, in the case of the Sunburst Arc we adopted zero attenuation for the UV lines, and used the attenuation from the JWST-only fit for the optical lines, as the latter is also consistent with the value obtained by \citet{Welch2025} from the Balmer decrements.
To derive temperatures and abundances via the $T_{\mathrm{e}}$-method, we used the \texttt{PyNeb} package, adopting the same transition proabilities and collisional excitation rates as in our photoionisation modelling, summarised in Table \ref{tab:atomic_parameters}. This ensures the best possible consistency between the abundances inferred through different approaches.
\begin{table*}
\centering
\caption{Atomic data.}
\begin{tabular}{lcc}
\toprule
\midrule
Ion & Transition Probabilities & Collision Strengths \\
\midrule
$\mathrm{C}^{2+}$ & \citet{Ferna2014} & \citet{Ferna2014} \\
$\mathrm{N}^{+}$  & \citet{Tachiev2001} & \citet{Tayal2011} \\
$\mathrm{N}^{2+}$ & \citet{Liang2012} & \citet{Liang2012} \\
$\mathrm{N}^{3+}$ & \citet{Ferna2017} & \citet{Ferna2017} \\
$\mathrm{O}^{+}$  & \citet{FroeseFischer2004} & \citet{Kisielius2009} \\
    $\mathrm{O}^{2+}$ & \citet{Tachiev2001}, \citet{Mao2020a} & \citet{Storey2014} \\
$\mathrm{S}^{+}$  & \citet{Kisielius2014} & \citet{Tayal2010} \\
$\mathrm{S}^{2+}$ & \citet{FroeseFischer2006} & \citet{Hudson2012} \\
$\mathrm{Ar}^{2+}$ & \citet{Wiese1969} & \citet{Galavis1995} \\
$\mathrm{Ne}^{2+}$ & \citet{McLaughlin2011} & \citet{McLaughlin2011} \\
\midrule
\bottomrule
\end{tabular}
\end{table*}
\label{tab:atomic_parameters}

The relative abundance of two ionic species X$^i$ and Y$^k$ is given by the ratio of the line intensities $I_\lambda$ and the emissivities $j_\lambda(n_\mathrm{e},T)$ as follows:
\begin{equation}
\frac{N\left(\mathrm{X}^i\right)}{N\left(\mathrm{Y}^{k}\right)}=\frac{I_{\lambda\left(\mathrm{X}^i\right)}}{I_{\lambda\left(\mathrm{Y}^k\right)}} \frac{j_{\lambda\left(\mathrm{Y}^k\right)}(n_\mathrm{e}, T)}{j_{\lambda\left(\mathrm{X}^i\right)}(n_\mathrm{e}, T)}.
\end{equation}
Below we summarise the steps followed for each element.

\textit{Oxygen.} We assumed that contributions from O and O$^{3+}$ are negligible, so that O/H $\approx$ O$^+$/H + O$^{++}$/H. We derived the abundances of O$^+$ and O$^{++}$ from [O II] $\lambda\lambda$3727,3729/H$\beta$ and [O III] $\lambda$5007/H$\beta$, respectively, assuming $T$(O III) and $T$(O II) for the high- and low-ionisation zones, and using the density derived from [S II] $\lambda\lambda$6717,6731.
To be consistent with the approach by \citet{Welch2025} for the Sunburst Arc we instead determined simultaneously $T$(S II) and $n_\mathrm{e}(\mathrm{S \ II})$, and used the former for the low-ionisation zone.
In our reference observations of RCJ2248-ID, [S II] $\lambda\lambda$6717,6731 lines are not detected. However, adopting the low-ionisation temperature and the [S II] $\lambda6731/\lambda6717$ ratio reported by \citet{Berg2025}, we would infer $n_\mathrm{e}(\mathrm{[S,II]}) = 2^{+5}_{-1} \times 10^3,\mathrm{cm}^{-3}$. 
For this analysis we used only UV diagnostics instead, following \citet{Topping2024}, as representative of common $T_\mathrm{e}$-approach when only UV density tracers are present. In particular, for this galaxy we adopted the density from N IV] $\lambda\lambda$1483,1486 for the high-ionisation zone and simultaneously derived $T$(O III) and $n_\mathrm{e}(\mathrm{C \ III]})$. For the low-ionisation zone, we fixed density value to 300 cm$^{-3}$ and adopted the temperature-temperature relation from \citet{Campbell1986}.

\textit{Carbon.} To compute carbon abundance we used C III] $\lambda\lambda$1907,1909/H$\beta$ as proxy for C$^{++}$, and we adopted $T$(O III) and $n_{\mathrm{e}}$(C III]). We then applied the ICF from \citet{Berg2019} to obtain log(C/O).
As sanity check, we also considered the empirical relation from \citet{PerezMontero2017}, which allows derivation of C/O from C III] $\lambda\lambda$1907,1907, C IV $\lambda\lambda$1548,1551, and O III] $\lambda\lambda$1660,1666. In both the RXCJ2248-ID and the Sunburst Arc, the two estimates are in good agreement.

\textit{Nitrogen from optical lines.} Since [N II] $\lambda$5755 is not detected in any of our sources, we derived the nitrogen abundance from [N II] $\lambda$6583/[O II] $\lambda\lambda$3727,3729, adopting $T$(O II) (or $T$(S II) for the Sunburst Arc), and applying the ICF from \citet{Amayo2021}.

\textit{Nitrogen from UV lines.} In RXCJ2248-ID, both N IV] and N III] muliplets are detected. We derived N$^{++}$/H and N$^{+++}$/H from those transition assuming the density from N IV] $\lambda\lambda$1483,1486 and $T$(O III). Following \citet{Topping2024}, we used no ICF to compute the total N/O.
To compare various methods, we also computed N/O from C/O and the ratio C$^{++}$/N$^{++}$, given the similarities in the two ionisation potentials \citep{Berg2019}.
Since in the Sunburst Arc only N III] $\lambda\lambda$1750,1753 is detected, we computed N/O from UV lines using the latter method. 

\textit{Neon.} We derived Ne$^{++}$/H from [Ne III] $\lambda$3869/H$\beta$ using $T$(O III), and the ICF by \citet{Dors2013} to obtain log(Ne/O). 

\textit{Sulfur.} We computed S$^+$/H and S$^{++}$/H from [S II] $\lambda\lambda$6717,6730/H$\beta$ and [S III] $\lambda$9069/H$\beta$, respectively, using $T$(OIII) and $T$(OII). Then we applied the ICF by \citet{Izotov2006} to obtain total S/O.
In M~4327 and the Sunburst Arc, both the auroral ([S II] $\lambda\lambda$4068,4076 and [S III] $\lambda$6312) and nebular lines are detected at S/N > 3. This allows a direct measurement of the sulfur abundance using both $T$(S II) and $T$(S III). In this case, we find $\log(\mathrm{S/O})$ values that are $\simeq0.2 - 0.3$ dex higher than those obtained using oxygen temperatures, but still consistent within uncertainties.  
We adopted temperatures from oxygen, as they yield more precise abundance estimates (since are based on higher S/N measurements).

\textit{Argon.} Since in M~4327 [Ar IV] lines are not detected, we estimated the argon abundance from [Ar III] $\lambda$7135/H$\beta$ using $T$(O III), and applied the ICF from \citet{Izotov2006}. Ar$^{++}$, with an ionisation potential of 27.6 eV, traces an intermediate-ionisation region between O$^{++}$ and S$^{++}$. 
\section{The HOMERUN models for the Sunburst Arc}
\label{app:B}
In this Appendix, we discuss more in detail the HOMERUN models of the Sunburst Arc presented in Sec. \ref{sec: Sunburst fiducial model} (Appendix \ref{app:B1}). We also qualitatively comment on how including VMS in our stellar continua could affect our results (Appendix \ref{app:B2}).

\subsection{Model assumption and agreement with observations}
\label{app:B1}
\begin{figure}[ht]
    \centering
    \includegraphics[width=\linewidth]{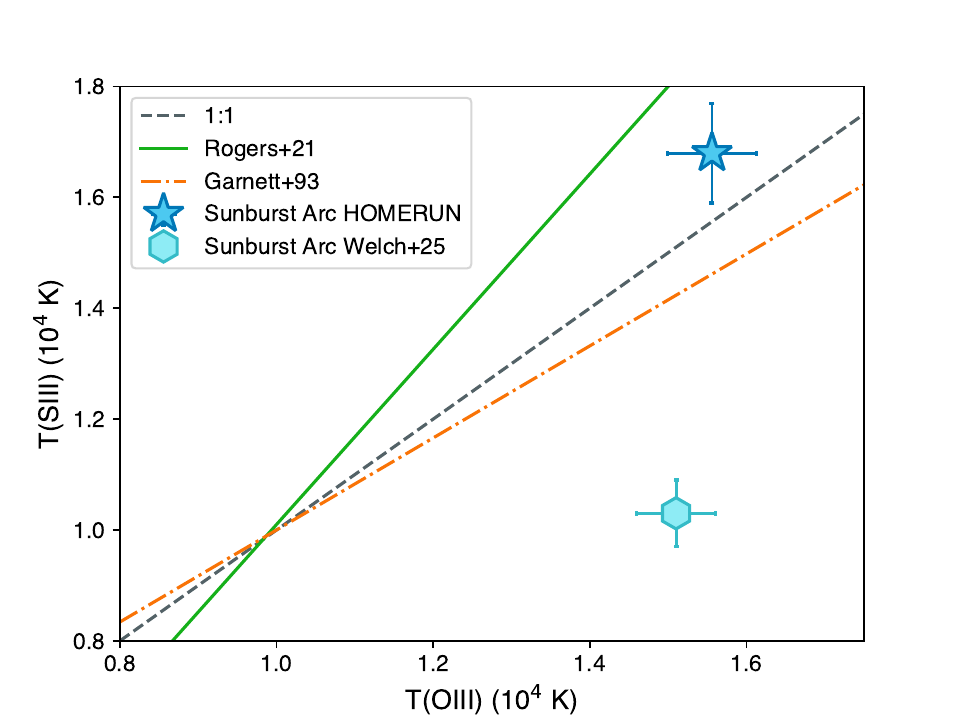}
    \caption{Relation between the temperature from [S III] and that from [O III] obtained when using [S III] $\lambda$6312 predicted from the best-fit model (star), and from observations (hexagon). For reference, also identity (grey dashed line), and analytical temperature-temperature relations from \citet{Garrnett92} (solid green line) and \citet{Rogers2021} (orange dot-dashed line) are shown.}
    \label{TSIII Sunburst}
\end{figure}

Despite the importance of auroral lines in constraining ISM temperature, we chose to exclude [S III] $\lambda$6312 from our HOMERUN fits. This conservative decision is motivated by the unexpectedly low $T$(S III) derived from sulfur lines. As reported by \citet{Welch2025}, the $T$(S III) value is significantly lower than both $T$(O III) and predictions from temperature–temperature relations. Our best-fit fiducial model also hints towards a higher temperature, as shown in Figure \ref{TSIII Sunburst}.
Moreover, \citet{Stanton2025a} argue that assuming $T$(O III) instead of $T$(S III) when computing the argon abundance yields lower $\log$(Ar/O) values, more consistent with their galaxy sample and predictions from models. These authors conclude that the [S III] $\lambda$6312 measurement might be unreliable.

\begin{figure}[ht]
    \centering
    \includegraphics[width=\linewidth]{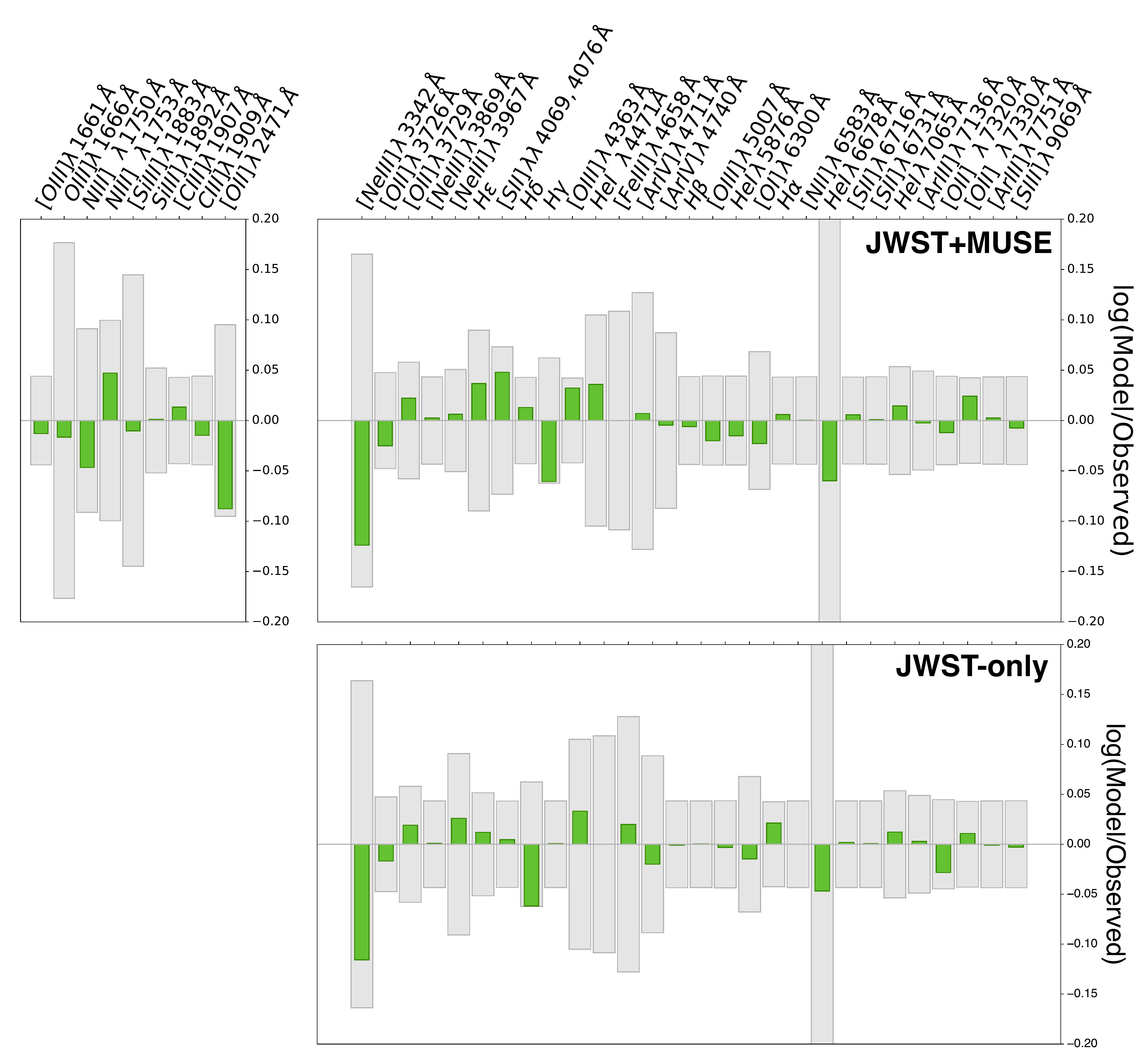}
    \caption{Similar to the right panel of Figure \ref{M4327 chisq}, but for our fiducial JWST-only (bottom) and JWST+MUSE (top) fit of the Sunburst Arc. In this case, we show the agreement between model and observations with green ($<1\sigma$), orange ($<2\sigma$), orange ($<3\sigma$) or red bars, while grey bars represent the acceptable discrpancey (i.e., $\sigma$) The left block report lines from MUSE \citep{Pascale2023}, and the right one those from JWST \citep{Welch2025}}.
    \label{Sunburst fiducial fits lines}
\end{figure}

\begin{figure}[ht]
    \centering
    \includegraphics[width=\linewidth]{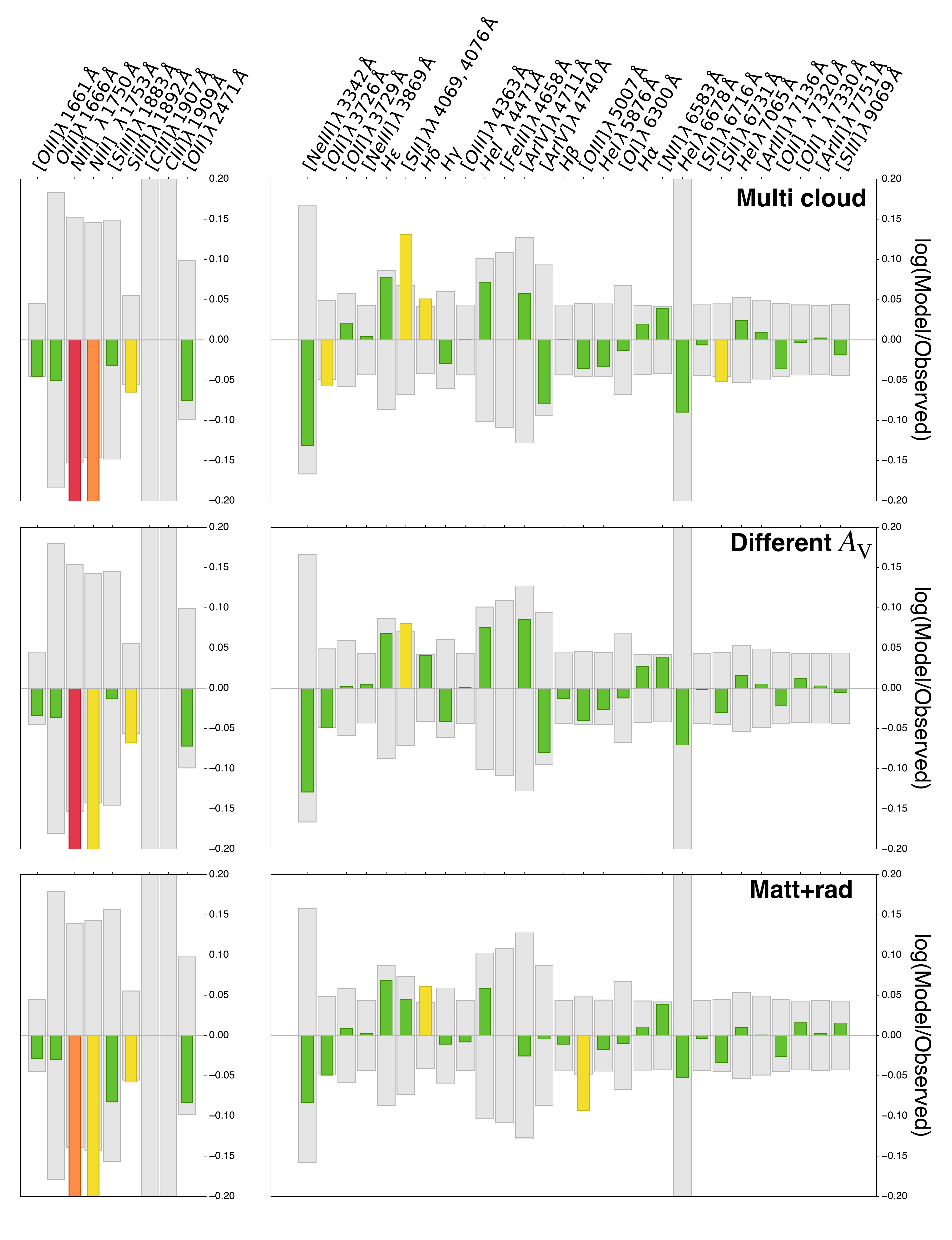}
    \caption{Analogous to Figure \ref{Sunburst fiducial fits lines}, but showing the agreement with observations for all the unsuccessful models described in Section \ref{sec: Sunburst fiducial model}. 
    } 
    \label{Sunburst bad fits lines}
\end{figure}

Figures \ref{Sunburst fiducial fits lines} and \ref{Sunburst bad fits lines} report the agreement between observed emission lines and model predictions of HOMERUN best fits, for both the fiducial and unsuccessful models described in Sec. \ref{Sunburst models}.
In particular, Fig. \ref{Sunburst bad fits lines} further strengthens the conclusion of Sec. \ref{sec: Sunburst fiducial model}: all unsuccessful models reproduce the emission lines included in the HOMERUN fit typically within the assumed error (or within $2\sigma$), but cannot explain nitrogen optical and UV features simultaneously.
However, limiting the fit to optical transitions or including chemical stratification allows the best-fit models to predict the observed fluxes within their error (Fig. \ref{Sunburst fiducial fits lines}).

\subsection{Very massive stars}
\label{app:B2}
Prior to JWST, several studies, and particularly \citet{Mevs2023} invoked the presence of VMS winds in the Sunburst Arc to explain the strong and broad He II $\lambda$1640 emission.
More recent works based on simulations and chemical evolution models, also required SMS or VMS to reproduce the observed abundance patterns in many N-emitters \citep[e.g.,][]{Charbonnel2023, MarquesChaves2024, Tapia2024, Shi2025}.
While our ionising spectra include stars with masses up to 300 M$_\odot$, BPASS does not fully capture the evolution and atmospheres of VMS.
In particular, stellar winds in BPASS are treated similarly for massive and VMS \citep{Schaerer2025, Upadhyaya2024}, whereas \citet{Mevs2023} used dedicated models for VMS. 

Considering the models by \citet{Martins2022} with stellar metallicity of $0.4\ \mathrm{Z}_{\odot}$ (i.e., $\log(Z_{\star})=-2.3$) and age 1 Myr, \citet{Schaerer2025} also noted that the spectral energy distribution remains nearly unchanged below 35 eV in the presence or absence of VMS.
Therefore, in such case, the inclusion of VMS does not affect optical line ratios, which lie mostly below this threshold.
In the same models, the VMS spectra are actually softer in the $34-54$ eV range and at higher energies, where most ionising photons are absorbed within the stellar atmosphere and dense stellar winds.
\citet{Martins2025} extended their previous models to lower stellar metallicities and different ages. 
These authors find that the presence of VMS indeed hardens the spectra as $Z_\star$ decreases ($0.2,\ 0.1,\ 0.01\ \mathrm{Z}_{\odot}$).
However, \citet{Martins2025} also warn about the possible uncertainties inherent in generalising their results: in general, the actual shape of the ionising spectrum depends on the assumed mass-loss rate in stellar winds, stellar ages, star formation histories, and energies, since the physics involved at such high energies is particularly challenging to model. As a consequence, observational constraints on VMS evolution are crucial to better characterise these regimes.

The single-cloud models considered in HOMERUN are build with a rather coarse grid in stellar metallicity (Table \ref{tab:models}). 
Moreover, HOMERUN is not yet optimised to recover the exact shape of the stellar continuum and thus not designed to constrain stellar population properties \citep{Marconi2024}.
Therefore, within our approach we cannot confirm nor confute the presence of VMS or SMS in the considered systems, but just assess that overall the ionisation properties are similar to those of the best-fit continuum. However, the presence of such peculiar stellar populations could change both the inferred best-fit age and the stellar metallicity. For instance, we also cannot exclude that the very-low stellar metallicity ($\log(Z_{\star})=-4$) found for the best-fit of RXCJ2248-ID might also be explained with a higher $Z_\star$ but including VMS/SMS or harder ionising stellar sources.

\section{The HOMERUN models for the RXCJ2248-ID}
\label{app:AppC}
As in Appendix \ref{app:B}, we report below the performance of our best-fit models across the full set of emission lines considered (Appendix \ref{app:AppC1}), together with a few considerations on how atomic data may influence N/O estimates derived from UV lines (Appendix \ref{app:C3}).

\subsection{Model agreement with observations}
\label{app:AppC1}
\begin{figure}[ht]
    \centering
    \includegraphics[width=\linewidth]{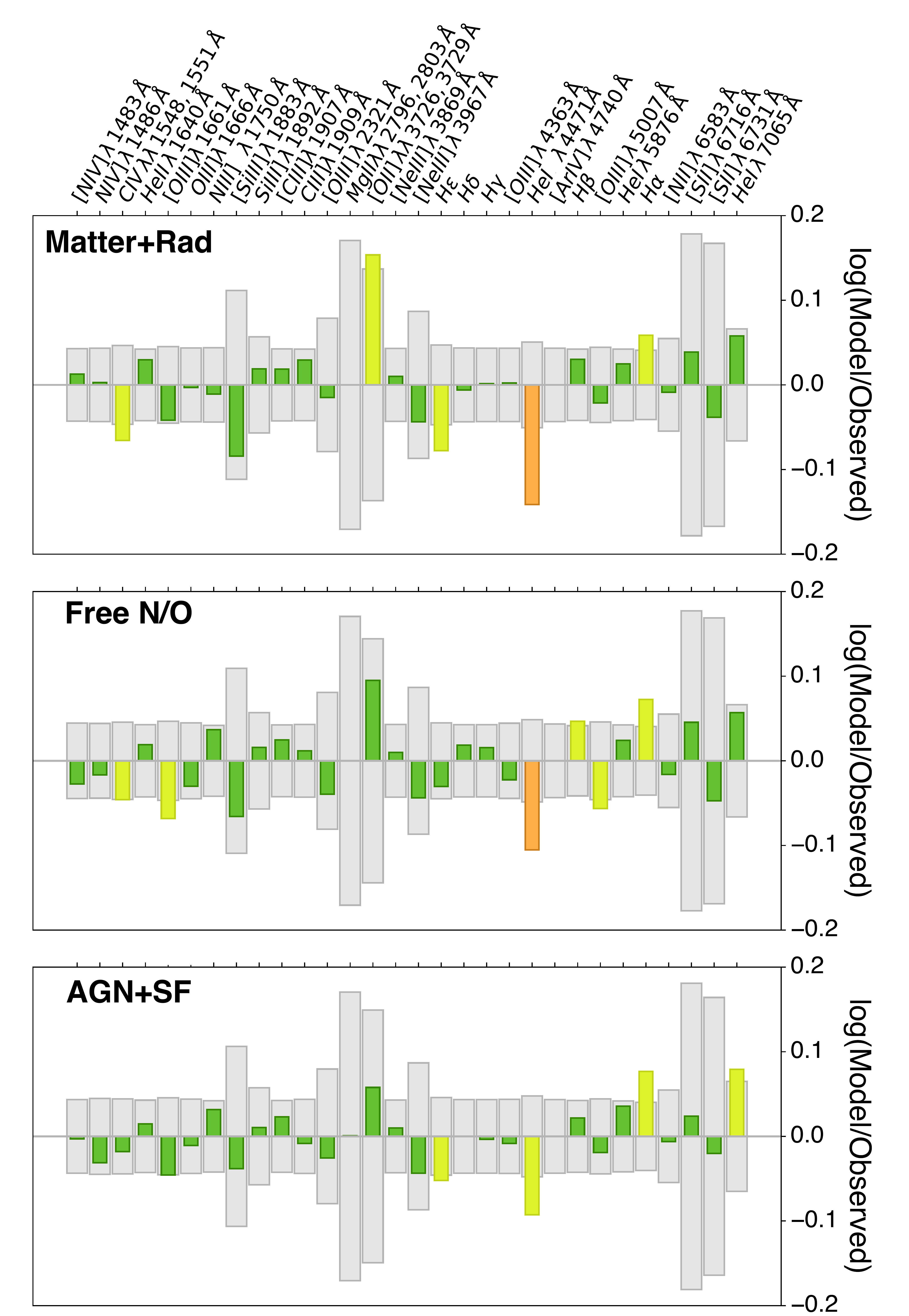}
    \caption{Same as Figure \ref{Sunburst fiducial fits lines}, but for RXCJ2248-ID best-fit models presented in Sec. \ref{sec: Topping multi comp mods}.}
    \label{Topping lines}
\end{figure}

Figure \ref{Topping lines} shows the agreement between the observations and the HOMERUN best-fit predictions for the models presented in Sec. \ref{sec:Topping_models}.
These models are able to reproduce all emission lines generally within the assumed uncertainties over the entire wavelength range.
A notable exception is the He I $\lambda4471$ multiplet. However, this emission feature is only slightly underpredicted by our AGN+SF model, and remains within $<2\sigma$ for the two star-forming models. As a sanity check, we verified that excluding this line from the fit does not affect the derived abundances within their uncertainties.
In general, the best-fit model values for He I $\lambda4471$/H$\beta$ (and for H$\alpha$) are consistent with the measurements reported by \citet{Berg2025}.
While the AGN+SF model provides the best overall agreement with the observations, we decided to adopt the ‘Matter+Rad’ model as our fiducial one, given the detection of WR features in this galaxy (see Sec. \ref{sec:Topping_models}).

\subsection{Possible impact of atomic data on N/O from UV lines}
\label{app:C3}
In our analysis of RXCJ2248-ID the N/O values derived considering UV lines are typically $\simeq0.1-0.15$ dex lower than those reported in the literature \citep{Topping2024, Berg2025, Isobe2025, Martinez2025}, though still marginally consistent within the uncertainties. This discrepancy primarily arises from the different atomic data assumed in our work, and in particular from the choice of collisional strength rates for N IV, which differ from the default values in \texttt{PyNeb} (Tab. \ref{tab:atomic_parameters}). Indeed, adopting the latter (while keeping all other atomic data from CLOUDY), we obtain a $\log\mathrm{(N/O)}=-0.43$, in better agreement with published values. Using the collisional strengths from \citet{Ferna2014} would lead to an even higher $\log\mathrm{(N/O)}=-0.29$.
N IV is well known to be a problematic ion, and there is no general consensus on which atomic data should be preferred \citep[e.g.,][]{Aggarwal2017, DelZanna2019}. For example, \citet{DelZanna2019} showed that the intensities of the main N IV emission lines can vary by up to $20\%$ depending on the method used to derive the atomic data. However, the same authors also argue that such differences should only mildly affect plasma diagnostics at the relatively low densities typical of gaseous nebulae (i.e., $n_\mathrm{e} < 10^{15}$ cm$^{-3}$).
While a detailed assessment of the impact of different atomic datasets and a systematic comparison of the methods used to derive them are both far beyond the scope of this paper, we caution that N/O abundances derived from N IV UV lines might also be sensitive to the specific atomic data considered.

\section{Other element abundances}
\label{app:AppD}

\begin{table*}
    \centering
    \caption{HOMERUN abundance ratios not discussed in the main text for the Sunburst Arc and RXCJ2248-ID.}
    \label{tab:abundances_appendix}
    \renewcommand{\arraystretch}{1.3} 
    \begin{tabular}{lcccccc}
    \toprule
    \midrule
    & \multicolumn{2}{c}{The Sunburst Arc} 
    & \multicolumn{3}{c}{RXCJ2248-ID} 
    & Sun$^a$ \\
    & JWST-only 
    & \shortstack{Fiducial\\JWST+MUSE} 
    & \shortstack{Fiducial\\Matter+Rad} 
    & AGN+SF 
    & Free N/O 
    & \\
    \midrule
    $12+\log(\mathrm{He}/\mathrm{H})$ 
    & $10.95^{+0.06}_{-0.03}$ 
    & $11.08^{+0.02}_{-0.06}$ 
    & $11.09^{+0.09}_{-0.04}$ 
    & $11.18^{+0.01}_{-0.06}$ 
    & $11.09^{+0.07}_{-0.03}$ 
    & $10.914 \pm 0.013$ \\

    $\log(\mathrm{C}/\mathrm{O})^b$  
    & -- 
    & $-0.556^{+0.128}_{-0.003}$ 
    & $-0.465^{+0.011}_{-0.029}$ 
    & $-0.50^{+0.07}_{-0.05}$ 
    & $-0.53^{+0.05}_{-0.05}$ 
    & $-0.23\pm0.08$ \\

    $\log(\mathrm{Ne}/\mathrm{O})$ 
    & $-0.58^{+0.04}_{-0.04}$ 
    & $-0.55^{+0.02}_{-0.03}$ 
    & $-0.62^{+0.09}_{-0.04}$ 
    & $-0.61^{+0.06}_{-0.07}$ 
    & $-0.63^{+0.07}_{-0.04}$ 
    & $-0.63\pm0.09$ \\

    $\log(\mathrm{Mg}/\mathrm{O})$ 
    & -- 
    & -- 
    & $-1.45^{+0.05}_{-0.45}$ 
    & $-1.57^{+0.07}_{-0.40}$ 
    & $-1.57^{+0.07}_{-0.40}$ 
    & $-1.14\pm0.07$ \\

    $\log(\mathrm{Si}/\mathrm{O})$ 
    & -- 
    & $-1.76^{+0.27}_{-0.02}$ 
    & $-1.44^{+0.11}_{-0.04}$ 
    & $-1.45^{+0.05}_{-0.15}$ 
    & $-1.54^{+0.19}_{-0.07}$ 
    & $-1.18\pm0.07$ \\

    $\log(\mathrm{S}/\mathrm{O})$  
    & $-1.77^{+0.02}_{-0.25}$ 
    & $-1.88^{+0.05}_{-0.04}$ 
    & $-1.55^{+0.06}_{-0.02}$ 
    & $-1.68^{+0.04}_{-0.16}$ 
    & $-1.48^{+0.05}_{-0.07}$
    & $-1.57\pm0.07$ \\

    $\log(\mathrm{Ar}/\mathrm{O})^c$ 
    & $-2.43^{+0.03}_{-0.23}$ 
    & $-2.60^{+0.07}_{-0.04}$ 
    & $-2.25^{+0.06}_{-0.07}$ 
    & $-2.27^{+0.06}_{-0.08}$ 
    & $-2.23^{+0.04}_{-0.9}$
    & $-2.31\pm0.14$ \\

    $\log(\mathrm{Fe}/\mathrm{O})$ 
    & $-2.04^{+0.05}_{-0.15}$ 
    & $-1.99^{+0.03}_{-0.05}$ 
    & -- 
    & -- 
    & -- 
    & $-1.23\pm0.08$ \\
    \midrule
    \bottomrule

\end{tabular}
\tablefoot{$^{(a)}$ Solar values from \citet{Asplund2021}.
$^{(b)}$ For reference, we also report the log(C/O) expected from the relation by \citet{Nicholls2017} at our fiducial-model metallicities: for the Sunburst Arc $\log\mathrm{(C/O)}=-0.62$, and for RXCJ2248-ID $\log\mathrm{(C/O)}=-0.72$.
$^{(c)}$ For RXCJ2248-ID, argon abundances are determined by fitting the sole [Ar IV] $\lambda$4711/H$\beta$ ratio, taken from \citet{Berg2025} since argon transitions are not detected in \citet{Topping2024} (also Sec. \ref{sec:Topping_models}). As a consequence, we caution that this value might be particularly sensitive to possible differences in the two datasets, which we do not consider in the current method for estimating uncertainties.  }

\end{table*}

So far, we limited our analysis of chemical abundances in the Sunburst Arc and RXCJ2248-ID to oxygen and nitrogen. However, investigating relative abundances of elements with different nucleosynthetic origins can provide valuable insights into chemical enrichment processes \citep[e.g.,][]{Kobayashi2025}. A detailed analysis will be presented in future works, but for completeness we report in Table \ref{tab:abundances_appendix} the best-fit model abundances for all the elements not discussed before, along with a few considerations below.

\textit{Helium.} Previous studies suggested a correlation between nitrogen and helium enrichment. If nitrogen excess originates from CNO processing in massive stars, and hydrogen is simultaneously burned into helium in outer layers, then stellar winds may eject both into the surrounding ISM, producing regions enriched in both N and He. This is consistent with observations of second-generation stars in globular clusters, which often show helium enrichment, making helium a possible tracer of clustering efficiency \citep[e.g.,][and references therein]{Ji2025a}.

Intriguingly, if we also allow helium abundance to vary freely in the fit of RXCJ2248-ID, we found that the nitrogen-rich component shows a helium abundance approximately twice the solar value, while the other component have normal levels of both N and He.
However, leaving He abundance free does not significantly improve the fit quality, and is not required, but only represents an additional degree of freedom.

In general, \citet{Ji2025a} do not find a clear trend of He enrichment in N-rich galaxies within their sample.
High He abundance can be mimicked by assuming emission predominantly arises from very-dense clouds ($n_\mathrm{e} \sim 10^5$\,cm$^{-3}$), where collisional excitation significantly contributes to He line fluxes \citep[e.g.,][]{Thuan1996,Izotov2014,Yanagisawa2024, Katz2024, Ji2025a, Berg2025}.
While HOMERUN self-consisently considers such high densities, our best-fit models for both the Sunburst Arc and RXCJ2248-ID (regardless of the adopted assumptions) yield high helium abundances compared to the solar value $12+\log(\mathrm{He/H})=10.914 \pm 0.013$ \citep{Asplund2021} (Table \ref{tab:abundances_appendix}).
For RXCJ2248-ID, we find $11.09^{+0.09}_{-0.04}$ for the 'Matter+Rad' model. We also tested excluding He I $\lambda$4471, which is slightly underpredicted (still within $3\sigma$), or He II $\lambda$1640, but obtained consistent He abundances in both cases.

\textit{Carbon.} 
Carbon is produced through multiple channels, including massive and intermediate-mass stars, and is therefore sensitive to star formation history, enrichment timescales, and nucleosynthetic processes.
In addition, commonly used UV diagnostic diagrams at high redshift involve carbon emission lines, which in turn depend on the C/O ratio \citep[e.g.,][]{Gutkin2016,Hirschmann2019}.
While the best-fit $\log(\mathrm{C/O})$ for the Sunburst Arc is consistent with the estimate by \citet{Pascale2023}, the model-derived $\log(\mathrm{C/O})$ abundance ratio in RXCJ2248-ID is $\simeq 0.15$–$0.25$ dex higher than the values reported by \citet{Topping2024}, \citet{Isobe2025}, and also \citet{Berg2025}, though the latter used different flux measurements. In contrast, our best-fit model C/O is more in agreement with the estimate from the calibration of \citet{PerezMontero2017} ($\log(\mathrm{C}/\mathrm{O}) = -0.62 \pm 0.14$), as well as with the $T_{\mathrm{e}}$-based value obtained using the same atomic parameters adopted in CLOUDY ($\log(\mathrm{C}/\mathrm{O}) = -0.60 \pm 0.10$).
As for the N IV (Appendix \ref{app:C3}), a full treatment of the impact of different atomic datasets on C/O determination is beyond the scope of this work.

\citet{Gutkin2016} pointed out that non-solar C/O ratios can affect observed line ratios, especially in UV diagnostics. As discussed in Sec. \ref{subsec:grids of cloudy models} and in \citet{Marconi2024}, our single-cloud models do not explicitly include variations with respect to the \citet{Nicholls2017} C/O-O/H scaling in the photoionisation calculations. Even though this effect might play a non-negligible role in shaping the observed emission, our fiducial best-fit models reproduce most emission lines within errors (or within 10\%) without requiring significant deviations from the adopted scaling relation. Still, we plan to explore this aspect in future work by expanding our modelling approach.

\textit{Silicon.} 
Silicon is mainly produced by massive stars, with a minor contribution from AGB stars and Type Ia supernovae \citep[e.g.,][]{Kobayashi2025}. Since silicon is a refractory species, its gas-phase abundance is strongly affected by dust depletion. As a consequence, Si/O is also a sensitive tracer of dust content and grain growth in the ISM.
The Si/O ratios reported in Table~\ref{tab:abundances_appendix} are sub-solar in both the Suburst Arc and RXCJ2248-ID. This values are in agreement with the presence of non-negligible dust content in both galaxies \citep[e.g.,][]{Pascale2023}, and consistent with the presence of dust grains in which silicon is more heavily depleted than oxygen \citep{Isobe2025}.  

\section{Correcting for temperature fluctuations}\label{app:J}

To address the issue of temperature fluctuations both observational- and simulative-based prescriptions have been developed. While HOMEURN allows a self-consistent treatment of ISM complexity, here we briefly discuss the approaches by \citet{Me2023a} and \citet{Cameron2023}.

Recently, \citet{Me2023a} proposed an empirical calibration to recover the correct metallicity when the right combination of diagnostics is available.
As an example, we follow their prescription for M~4327. Specifically, we extract the flux of [N II] $\lambda5755$  from the HOMERUN fiducial model described in Sec. \ref{sec:marta_HR} (this line is undetected in the JWST data). We use this model-predicted value to derive both $T$(N II) and $n_{\mathrm{e}}$ using \texttt{getCrossTemDen}, adopting [O II] $\lambda\lambda3727,3729/\lambda\lambda7323,7332$ as a density-sensitive diagnostic over the $n_\mathrm{e} \sim 10^3$–$10^6$ cm$^{-3}$ range \citep{Me2023}.
We obtain $n_\mathrm{e} = 682$ cm$^{-3}$ and $T$(N II) = 9903 K.
We then follow \citet{Me2023a} to account for temperature fluctuations and adopt their relation
$12 + \log(\mathrm{O}/\mathrm{H}) = -1.19 \times 10^{-4} \ T\mathrm{(N\,II)} + 9.68$,
which was calibrated on a sample of galactic and extragalactic HII regions, and ring nebulae.
With this correction we obtain $12 + \log(\mathrm{O}/\mathrm{H}) = 8.50$, in agreement with the best-fit metallicity from our fiducial HOMERUN model (Table \ref{tab:Marta properties}).

\citet{Cameron2023} used high-resolution radiation-hydrodynamic simulations of a dwarf galaxy to calibrate linear relations between the auroral-to-nebular line ratios and actual gas temperatures for a given transition. However, their calibrations are only valid for densities $\lesssim 10^3$ cm$^{-3}$, and therefore should not be directly applicable to high-redshift systems, where denser environments are common (e.g., Sec. \ref{sec:introduction}).

As a qualitative test, we follow a similar approach to that of the simulation, computing $T$([O III] $\lambda$5007) and $T$([O III] $\lambda$4363) from our grid of single-cloud models, weighting each by the respective emitted line flux. We find differences of $\sim1000 - 2000$ K between the two temperatures, in agreement with the $t^2$ estimates reported in Sec. \ref{sec:limitations of opt diagn}.

\begin{figure}[ht]
    \centering
    \includegraphics[width=\linewidth]{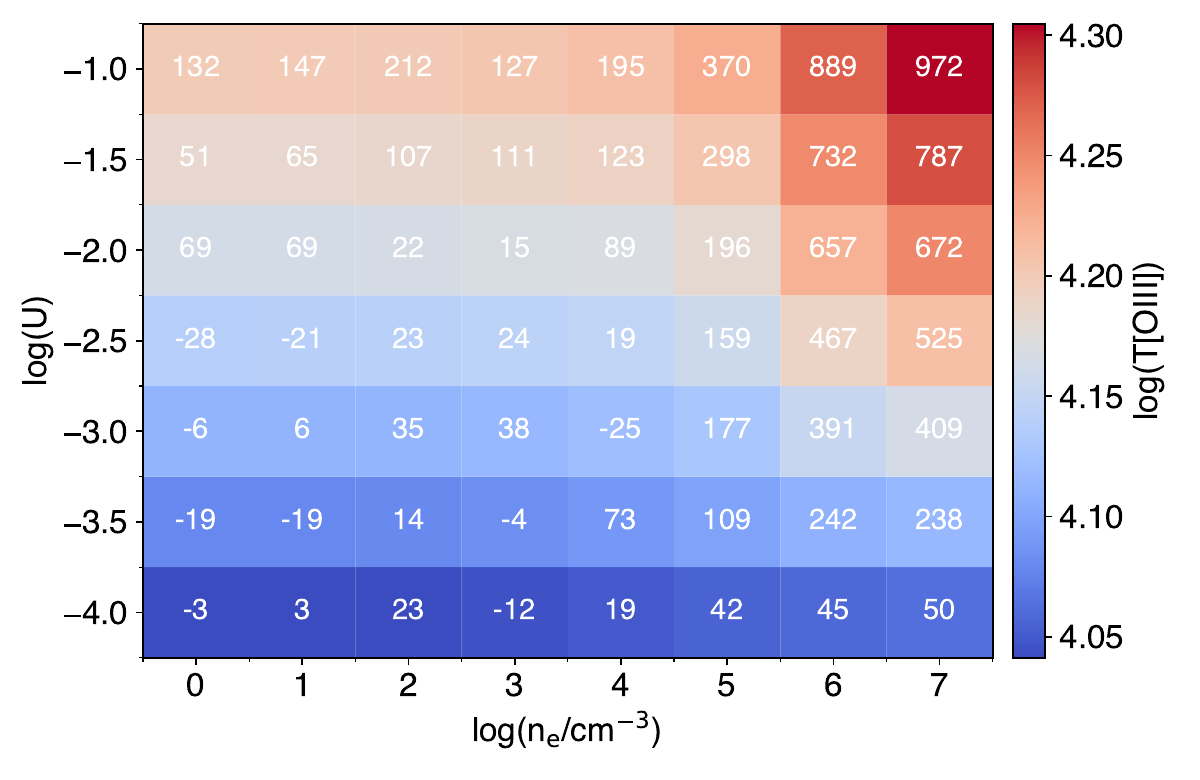}
    \caption{Temperature variations across our single-cloud model grid for a reference continuum with stellar age 1 Myr, stellar metallicity $\log(Z_{\star})=-2.7$ and gas-phase metallicity 12+log(O/H$)=7.89$. The colour scale represents the temperature derived from [O III] $\lambda$4363/[O III] $\lambda$5007, while in each cell we report the difference between the latter and the temperature of the O$^{++}$ zone extracted from CLOUDY output.}
    \label{T fluct}
\end{figure}

Lastly, we also consider our assumption of negligible temperature fluctuations within each single-cloud model (Sec. \ref{sec:limitations of opt diagn}).
Applying \citet{Cameron2023} corrections to models with $n_\mathrm{e} \leq 10^3$ cm$^{-3}$, we find that $T$([O III] $\lambda$5007) can be up to 3000 K lower than $T$ inferred from [O III] $\lambda4363$ / [O III] $\lambda5007$ ratio. The largest discrepancies occur in models with high ionisation parameters, in agreement with previous findings by \citet{Me2023a}.
In Figure \ref{T fluct}, we show $T$(OIII) from a reference grid of single-cloud models, alongside the difference with the O$^{++}$ zone temperature $T$(O$^{++}$) output by CLOUDY. The two estimates agree under "normal" conditions (log($n_{\mathrm{e}})\leq4$ and log($U)<-2$), but begin to diverge at high densities and ionisation parameters, although the mismatch remains moderate.
Therefore, even when the gas density is known, the internal temperature structure must be accounted for, especially in highly ionised and dense regions.

\end{appendix}

\end{document}